\documentclass[twocolumn,10pt]{article}
\usepackage{usenix2019_v3,epsfig,endnotes}
\pdfoutput=1

\pdfoutput=1
\usepackage[english]{babel}
\usepackage[colorinlistoftodos]{todonotes}

\usepackage{booktabs,caption}
\usepackage[flushleft]{threeparttable}
\usepackage{lstlinebgrd}
\usepackage{booktabs}
\usepackage{tabularx}
\usepackage{cite}
\usepackage{enumitem}
\usepackage{longtable,color,psfrag,dsfont,supertabular,bm,units}
\usepackage{amssymb,amsmath}
\usepackage{graphicx} 
\usepackage{threeparttable}
\usepackage{microtype}
\usepackage{verbatim}
\usepackage{xcolor, colortbl}
\usepackage{algorithmic}
\usepackage{multirow}
\usepackage{subfigure}
\usepackage{fullpage}
\usepackage{lscape}
\usepackage{afterpage}
\usepackage{tablefootnote, threeparttable}
\usepackage{comment}
\usepackage{cite}
\newcommand{\iotbench}{{\textsc{\small{IoTBench}}}\xspace}
\newcommand{\saint}{{\textsc{\small{SainT}}}\xspace}
\newcommand{\saints}{{\textsc{\scriptsize{SainT}}}\xspace}
\newcommand{\daint}{{\textsc{\small{IoTWatcH}}}\xspace}

\newcommand{\daints}{{\textsc{\scriptsize{IoTWatcH}}}\xspace}
\newcommand{\daintf}{{\textsc{\footnotesize{IoTWatcH}}}\xspace}
\newcommand{\smarthings}{{\textsc{\small{SmartThings}}}\xspace}
\setlist[itemize]{leftmargin=*}
\setlist[enumerate]{leftmargin=*}
 \usepackage[ruled,vlined,linesnumbered]{algorithm2e}
\usepackage{amssymb,amsmath}
\usepackage[hang,flushmargin]{footmisc}
\usepackage{lipsum}
\usepackage{xcolor}
\definecolor{light-gray}{gray}{0.90}

 \makeatletter
\newcommand{\algorithmfootnote}[2][\footnotesize]{%
 \let\old@algocf@finish\@algocf@finish% Store algorithm finish macro
 \def\@algocf@finish{\old@algocf@finish% Update finish macro to insert "footnote"
    \leavevmode\rlap{\begin{minipage}{\linewidth}
    #1#2
    \end{minipage}}%
  }%
}
\makeatother
\usepackage{geometry}
\iftrue
\newcommand{\leo}[1]{{\color{red}{\bf} #1}}
\else
\newcommand{\leo}[1]{}
\fi
\iftrue

\else
\newcommand{\leo}[1]{}
\fi

\iftrue
\newcommand{\berkay}[1]{{\color{blue}{\bf BC:} #1}}
\else
\newcommand{\berkay}[1]{}
\fi

\DeclareRobustCommand*\circled[1]{\tikz[baseline=(char.base)]{ \node[shape=circle,draw,color=white,fill=black,inner sep=0.5pt] (char){#1};}}

\usepackage{fancyvrb}
\usepackage[scaled=.8]{beramono}

\makeatletter 
\newcommand\semiHuge{\@setfontsize\semiHuge{22.72}{27.38}}
\makeatother 

\usepackage{listings}
\usepackage{color}
\newcommand{\bcircle}{
\begin{tikzpicture}
\filldraw[fill=black,draw=black] circle (2pt);
\end{tikzpicture}
}

\newcommand{\wcircle}{
\begin{tikzpicture}
\filldraw[fill=white,draw=black] circle (2pt);
\end{tikzpicture}
}

\def\eg{{e.g.},~}

\begin{document}
\title{{Real-time Analysis of Privacy-(un)aware IoT Applications}}
%\title{DaINT: Dynamic Multi-class Classification, Context, and Data Flow Semantics Analysis in Smart Home Applications}

%for single author (just remove % characters)
\author{
{\rm Leonardo Babun$^1$, Z. Berkay Celik$^2$, Patrick McDaniel$^3$, A. Selcuk Uluagac$^1$}\\ \\
\rm {$^1$Cyber-Physical Systems Security Lab, Florida Inernational University}\\ \rm{$^2$Department of Computer Science, Purdue University} \\ \rm{$^3$Systems and Internet Infrastructure Security Lab, Penn State University} \\
\rm {$^1$\{lbabu002, suluagac\}@fiu.edu, $^2$zcelik@purdue.edu, $^3$mcdaniel@cse.psu.edu}}

\begin{comment}
\author{Leonardo Babun}
\email{lbabu002@fiu.edu}
\affiliation{%
  \institution{Florida International University}
  \city{Miami}
  \state{Florida}
}

\author{Z. Berkay Celik}
\email{zcelik@purdue.edu}
\affiliation{%
  \institution{Purdue University}
  \city{West Lafayette}
  \state{Indiana}
}

\author{Patrick McDaniel}
\email{mcdaniel@cse.psu.edu}
\affiliation{%
  \institution{Penn State University}
  \city{State College}
  \state{Pennsylvania}
}

\author{A. Selcuk Uluagac}
\email{suluagac@fiu.edu}
\affiliation{%
  \institution{Florida International University}
  \city{Miami}
  \state{Florida}
}
\end{comment}

%for extended paper, optional text block \extended.
\newcommand{\showcomments}{O}% set to 0 to hide 
\newcommand{\comments}[1]{\ifthenelse{\equal{\showcomments}{1}}{#1}{}}

% make the title area
\maketitle

\begin{abstract}
%\boldmath
Users trust IoT apps to control and automate their smart devices. These apps necessarily have access to sensitive data to implement their functionality. However, users lack visibility into how their sensitive data is used (or leaked), and they often blindly trust the app developers. In this paper, we present \daint, a novel dynamic analysis tool that uncovers the privacy risks of IoT apps in real-time. We designed and built \daint based on an IoT privacy survey that considers the privacy needs of IoT users. \daint provides users with a simple interface to specify their privacy preferences with an IoT app. Then, in runtime, it analyzes both the data that is sent out of the IoT app and its recipients using Natural Language Processing (NLP) techniques. Moreover, \daint informs the users with its findings to make them aware of the privacy risks with the IoT app. We implemented \daint on real IoT applications. Specifically, we analyzed 540 IoT apps to train the NLP model and evaluate its effectiveness. \daint successfully classifies IoT app data sent to external parties to correct privacy labels with an average accuracy of 94.25\%, and flags IoT apps that leak privacy data to unauthorized parties. Finally, \daint yields minimal overhead to an IoT app's execution, on average 105 ms additional latency.   
\end{abstract}

\maketitle

\section{Introduction}
\label{sec:intro}
Users install IoT apps to manage and control smart devices such as smart thermostats, door locks, and cameras. These apps have access to sensitive information to implement their functionalities, communicate to external servers, and send notifications to users~\cite{celik2018program, Official, OpenHabGuideline, appleSecurity, sikder2018survey}. The sensitive information is either obtained through APIs provided by IoT platforms (e.g., whether the door is unlocked (\texttt{door.unlocked}) or the lights are off in the kitchen (\texttt{kitchenLights.off})); or they are simple texts defined by a developer or user at install-time, for instance ``the door is open'', and ``the kids returned to home''. Previous research works have demonstrated that IoT apps may leak sensitive information to unauthorized parties~\cite{saint, fernandes2016flowfence, sikder20176thsense, sikder2019context, privacymagazine}. Additionally, many IoT apps transmit data to remote servers without users' permission for data visualization and to profile user behaviors~\cite{magazine, acar2018peek}, such as their energy usage. However, users have no knowledge and control over what type of sensitive data the apps access or who else see these data~\cite{usb}. 

With the rapidly growing ecosystem of IoT devices and apps, users are oblivious to the privacy risks that IoT apps pose. Previous works focus on analyzing the app's context to extract permissions from IoT apps~\cite{jia2017contexiot, yuan, sikder2019aegis}. Other works use static analysis to find sensitive data-flows~\cite{saint} or inter-rule vulnerabilities in IoT deployments~\cite{iruler}. These approaches (albeit useful) have limitations in over-approximating data-leaks or in accessing real IoT data to perform their evaluations. A current dynamic method isolates sensitive data within sandboxes, which requires intensive developer effort~\cite{fernandes2016flowfence}. Mainly, none of the existing solutions consider sensitive data leaks of \textit{untainted strings} defined by a user or developer. In our investigations, we have found that out of 540 analyzed IoT apps, 64\% of them potentially leak sensitive data through simple strings. Lastly, current IoT systems do not consider the privacy preferences of users into their implementations. There exist effective privacy tools for the consumer in other domains such as mobile phones~\cite{flowcog, ArztFlowdroid, EnckTaintDroid, intro1, intro2, intro3}, yet these cannot be applied to analyze privacy risks of IoT apps as IoT apps possess unique challenges in terms of programming languages and structures.

In this paper, we introduce \daint, a novel dynamic analysis tool that uncovers the privacy risks of IoT apps at runtime. We designed and built \daint based on an IoT privacy survey of human subjects that use different IoT devices. The survey aimed at understanding the privacy concerns and expectations of users when they use IoT devices and apps. \daint provides users with a simple interface that allows them to specify their privacy preferences (\eg location, device states, etc.) at install-time. It then adds extra logic to the app's source code to collect app information at runtime. The collected information is used to classify the data sent out of the IoT app into privacy preferences of users through Natural Language Processing (NLP) techniques. Also, \daint analyzes the recipients of the data and detects leaks to unauthorized parties. Finally, \daint notifies the users about the sensitive data-leaks and privacy concerns in IoT apps, allowing them to make informed decisions about their privacy. Privacy concerns include any app behavior that would put the sensitive information at risk. For instance, an Internet communication that sends the sensitive information to a remote server in plain text.

To evaluate \daint, we trained an NLP model with IoT strings extracted from 380 SmartThings market apps. The model is used to classify IoT app strings (\eg ``the door is locked'', and ``kitchen lights are turned off'') to user privacy preferences. Then, we analyzed 160 different IoT apps to evaluate its accuracy at runtime. \daint successfully classified 146 IoT strings to privacy preferences with an average accuracy of 94.25\% and precision of 95\%. Among its findings, \daint also flags 35 IoT apps that leak sensitive data to unauthorized recipients. Finally, \daint yields minimal overhead to the IoT apps execution, introducing on average 105 ms additional latency. 

\vspace{0pt}\noindent\textbf{Summary of Contributions.} The contributions of this work are as follows:
\begin{itemize}
\itemsep0em 
\item We conducted an IoT privacy survey with 123 IoT users through a set of structured questions. Although the survey is not the major focus of this work, it is instrumental for the IoT community, researchers, and users to understand users' privacy concerns, their privacy preferences, and expectations when they use IoT devices and apps. 

\item We designed and built \daint\footnote{We made the \daintf freely available to the community at \texttt{\small{https://iotwatch.appspot.com/}}}, a dynamic privacy analysis tool for IoT apps. \daint adds extra capabilities to an IoT app's source code to provide users with a privacy interface and collect app data. The interface allows users to easily specify their privacy preferences such as device states and location. Also, it permits the monitoring of data leaks and privacy behaviors in IoT apps at runtime. Lastly, \daint informs the users when an app's leak matches with the privacy preference of a user.

\item We analyzed 540 current IoT apps during the implementation and evaluation of \daint. First, we trained \daint with 380 apps. Then we evaluated its accuracy on the remaining 160 IoT apps (120 market and 40 malicious IoT apps). \daint classifies 146 IoT strings that are sent out of the apps into privacy labels with 94.25\% accuracy. Additionally, it successfully flags 62 sensitive leakages (29 via messaging and 33 via Internet communication) to unauthorized parties in 35 IoT apps. \daint detects sensitive data leaks without significant overhead, introducing on average 105 ms latency to an app's execution.
\end{itemize}

\vspace{1pt}\noindent\textbf{Organization.} The rest of the paper is organized as follows. In Section \ref{sec:userstudy}, we summarize the results of our IoT privacy survey. In Section \ref{sec:problem}, we articulate the problem through a use case and present the threat model. In Sections \ref{sec:overview} and \ref{sec:architecture}, we introduce \daint's architecture and its main components. Implementation details of \daint are given in Section \ref{sec:implementation}. Further, we evaluate \daint in Section \ref{sec:evaluation} and, in Section \ref{sec:benefits}, we present some discussions. Lastly, we discuss the related work in Section \ref{sec:relatedwork}, and Section \ref{sec:conclusion} concludes the paper.
%\vspace{-0.1in}

%%%%%%%%%%%%%%%%%%%%%%%%%%%%%%%%%%%%%%%%%%%%%%%%%%%%%%%%%%%%%%%%%%%%%%%%%%%%%%%%%%%%%%%%%%%%%%%%

\section{IoT Privacy Survey}
\label{sec:userstudy}
We conducted an IoT privacy survey to understand the privacy concerns of IoT users when they use various IoT devices and apps. Although the survey is not the major focus of this work, it provides rich insights into the users experiences and expectations on control over IoT app permissions, IoT privacy nudges, and their interplay. The entire survey was authorized by the institutional ethics review board (IRB) and occurred between April 2019 and May 2019.

\vspace{1pt}\noindent\textbf{Privacy Survey Goals.} We aim to answer the following questions: (1) what are the privacy concerns of IoT users?, (2) is there a need for privacy analysis tools designed for IoT?, and (3) what are the user expectations, in terms of usability requirements, for privacy analysis tools? 

We created 26 different questions organized into three categories. These categories align with three specific privacy survey goals: (1) the characterization of the participants, (2) privacy concerns of IoT users, and (3) the need for IoT privacy analysis tools and their usability requirements. We provide the details of the questions in Appendix~\ref{sec:examplesurvey}) and present the profiles of participants and our key findings below.

\vspace{1pt}\noindent\textbf{Survey Overview and Recruitment.} We made the survey available to participants for four weeks. The users could access the survey and submit their responses via an online questionnaire hosted on Google Forms~\cite{googleforms}. The questionnaire included single choice questions (e.g., yes, no), multiple-choice questions, and free-form questions (detailed in Appendix~\ref{sec:examplesurvey}). We made all the questions required except for the ones requesting an additional explanation from users in the form of free-text input. Finally, we recruited the participants using recruitment emails sent to lists of students, faculty, and staff in our institutions. The emails included a brief explanation about the survey and link to the online form. 

\vspace{1pt}\noindent\textbf{Participant Characteristics.} 
We recruited 123 participants of which 69 participants (56.1\%) were in the range of 18-25 years old and 37 (30.1\%) were in the range of 26-35 years. The remaining 17 (13.8\%) participants were 36 years or older. The majority of the participants (110 (89\%)) had at least completed some bachelor-level courses and 37 (30\%) were enrolled in graduate-level courses. A total of 112 (91.05\%) users shared that they currently use or are planning to use IoT devices in their homes. Finally, 19 (15.4\%) participants knew how to develop their own IoT apps while 82 (66.7\%) participants had previous experience installing apps from an IoT market or via using the source code of IoT apps available online.

\vspace{1pt}\noindent\textbf{Ethics and Analysis.} The human subjects review board of our institutions approved the privacy survey. The participants had to be over 18 years old to participate. The survey did not collect any personal information from participants, other than an institutional email address that was requested for compensation purposes. We did not allow participants to submit multiple responses, but they had the chance to change their answers anytime before the survey closing date. We processed and accepted all the responses obtained from the participants. Further, we directly quantified the responses from single- and multiple-choice questions. Finally, we used two independent researchers to analyze the free-from responses and did not consider any answers flagged as potential outliers.

\vspace{1pt}\noindent\textbf{Compensation.} After the survey's closing date, every participant was compensated either with extra-credit in their coursework or a gift card with a monetary value. The student participants could opt for receiving extra credit or monetary compensation. Faculty and staff all received gift cards. 

%%%%%%%%%%%%%%%%%%%%
\subsection{Survey Results}
\vspace{1pt}\noindent\textbf{Privacy Concerns of IoT Users.}
The participants were concerned about their private information being inadvertently leaked to unauthorized parties. Specifically, 65 (52.8\%) participants felt uncomfortable about their personal data (\eg their password to login into edge devices), their behavior and habits (\eg when they go to sleep), location (\eg whether they are home or away), device's settings (\eg heating value of a thermostat) and time configuration (\eg when kids leave home), and device states (\eg whether the door is locked or not) being handled by IoT systems. Also, at least 89 (72.4\%) participants expressed to be aware of IoT apps collecting their sensitive information and sending it to remote servers for data analytics such as profiling their energy usage and for advertisement purposes~\cite{magazine}. Finally, 103 (83.7\%) participants expressed privacy concerns on the use of IoT systems, and 88 (71.5\%) mentioned having heard about privacy issues in IoT systems from the news or other media. 

\begin{table}[t!]
\centering
\normalsize
\setlength{\tabcolsep}{2.5pt}
\resizebox{\columnwidth}{!}{
\begin{tabular}{|lc|}
\hline 
{\begin{tabular}[c]{@{}c@{}}Expectations of survey's participants\end{tabular}} &
{\begin{tabular}[c]{@{}c@{}}\% Agreement\end{tabular}}\\ \hline
\hline
\rowcolor{light-gray}
{Real-time privacy analysis}  & 91.6\%  \\ 
{Configurable privacy preferences}  & 87.5\% \\ 
\rowcolor{light-gray}
{Control over unauthorized data disclosure}  & 86.6\% \\ 
{On-demand privacy controls}  & 81.6\% \\ 
\rowcolor{light-gray}
{Timely privacy notifications}  & 85.3\%\\ \hline \hline
\textbf{Inter-rater Reliability}  & \textbf{86.5\%}\\
\hline
\end{tabular}
}
\caption{Participant responses when asked about their expectations from a privacy analysis tool, which guided the design of \daint. The percentage of agreement among the survey's participants showed a strong inter-rater reliability.}
\vspace{-0.2in}
\label{tab:features}
\end{table}

%\subsection{Privacy Analysis Tools in IoT}\label{sec:tools}
\vspace{1pt}\noindent\textbf{The Need for Real-time Privacy Tools.}
In total, 112 (91.1\%) participants raised broad concerns about lack of an existing tool that informs the user regarding potential privacy risks of IoT systems in real-time. Also, 119 (96.74\%) participants found the idea of using a tool to uncover privacy risks in IoT highly plausible. Our participants were willing to use automatic tools that modify (i.e., instrument) original IoT apps to enable privacy analysis in real-time. Out of the 123 participants, 119 (96.74\%) expressed their support to this option if the tool is verified by the IoT platform.

%\subsection{User-friendly Features}\label{sec:features}
\vspace{1pt}\noindent\textbf{User's Expectations.}
To understand the characteristics of an easy-to-use privacy tool, we asked participants a set of questions to evaluate their usability expectations. Table~\ref{tab:features} summarizes the participants' responses. Here, we detail the percentage of agreement among users for every privacy feature and the inter-rate reliability score. Finally, we evaluated the participants' approval of four different privacy labels utilized to classify information accessed by IoT apps. The term ``Device-info'' was considered appropriate to define information from devices (e.g., device states or device type) by 109 (88.6\%) participants, while the label ``User-behavior'' received 103 (83.7\%) positive responses to define information related to the user (e.g., what the user does, how the user configure his/her IoT system). Furthermore, the label ``Location'' was approved by 110 (89.4\%) participants to define information related to the location of devices and users. Lastly, the label ``Date-time'' was approved in 100 (81.3\%) responses to define timing-related information.

\subsection{Summary of Findings}%\label{sec:summary}
Our findings shed new light on the need for cooperative privacy management practices between users and IoT markets, which mitigate privacy risks based on user privacy preferences. Table~\ref{tab:features} presents the needs of the users that align with their control over their privacy preferences. We obtained an inter-rate reliability of 86.5\%, which can be considered strong. Here, we summarize the privacy features that guides design and development of \daint considering participants responses. 

\vspace{1pt}\noindent\emph{Real-time Privacy Analysis.} 
The participants reflected their opinion about being aware of apps privacy behavior in what information leaves an IoT system and where it is transmitted. Additionally, they reported that they expect to have minimal configuration when new devices are dynamically plugged into their IoT systems and new IoT apps are installed. 

\vspace{1pt}\noindent\emph{Configurable Privacy Preferences.} The participants mentioned that fears about lack of privacy preference controls limit their willingness to use IoT devices. For instance, they prefer to have categories that define a high-level category that shows the information-type leaving the IoT system, such as ``the door is locked'' associated with a specific privacy label and ''the mode is changed to sleep`` with another label.

\vspace{1pt}\noindent\emph{Unauthorized Data Disclosure.} The participants like having better control over the disclosure of any private information. For instance, they prefer to be notified when IoT systems share their data with other parties. Additionally, participants mentioned strategies that notify unencrypted Internet requests or hard-coded messaging recipients, which mitigates the consequences of privacy violations.

\vspace{1pt}\noindent\emph{On-demand Privacy Controls and Privacy Notifications.} The majority of the participants acknowledged the effectiveness of configurable privacy preferences over the IoT apps. They mentioned these configurations help them have a better experience with a few numbers of notifications and minimal runtime delay.
 
%%%%%%%%%%%%%%%%%%%%%%%%%%%%%%%%%%%%%%%%%%%%%%%%%%%%%%%%%%%%%%%%%%%%%%%%%%%%%%%%%%%%%%%%%%%%%%%%%%%%%%%

\section{Problem Statement and Threat Model}
\label{sec:problem}

\begin{figure}[t!]
\centering
    {\includegraphics[width=0.4\textwidth]{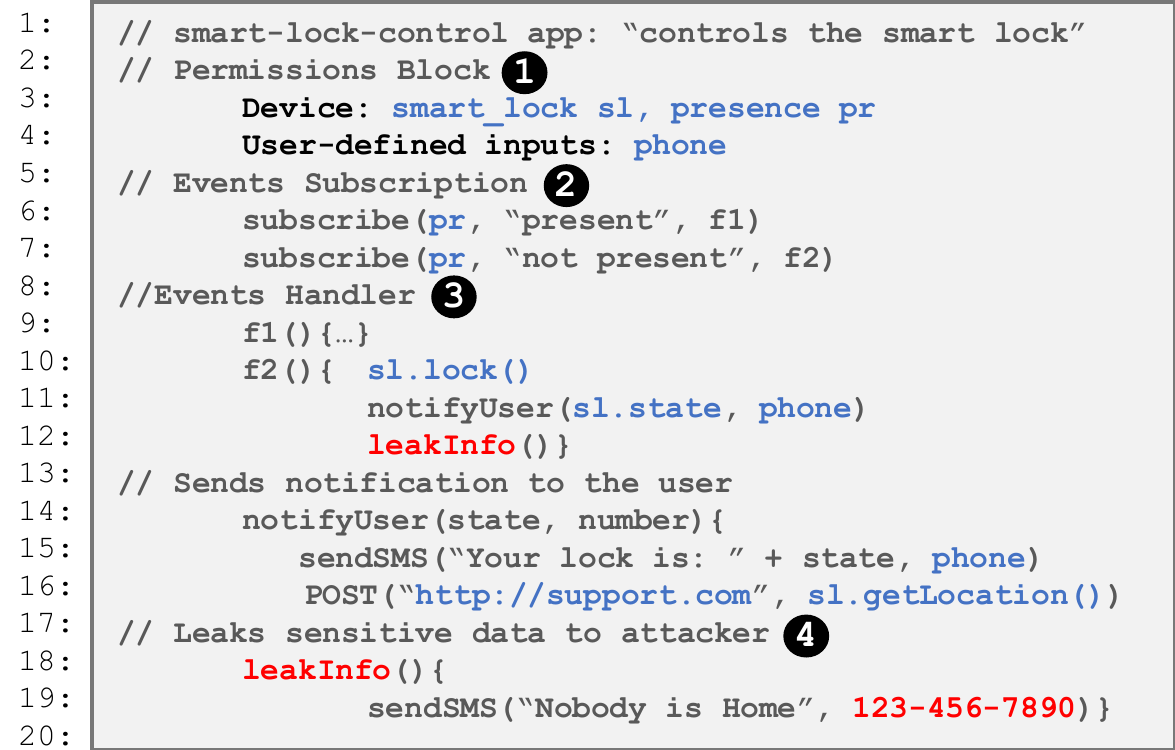}}
    \caption{An example IoT app leaking sensitive data to a hard-coded phone number and performing insecure HTTP calls.}
    \vspace{-0.2in}
    \label{fig:originalApp}
\end{figure}

\noindent\textbf{Problem Statement.} We use an example source code abstracted from a \texttt{smart-lock-control} app (Figure~\ref{fig:originalApp}) to illustrate the privacy concerns and behaviors in IoT apps. The expected behavior of the app is to lock the door and notify to user-defined contacts that the door is locked when the user leaves the house. At install-time, the user grants permissions to the smart lock, presence sensor, and enters a phone number for messaging notifications (\circled{1}). The app subscribes to two event handlers \texttt{f1} and \texttt{f2} to implement the app functionality. The event handlers are invoked based on the presence sensor's state (user-present and user-not-present) (\circled{2}). When the user leaves home, ``not-present'' event handler (i.e., \texttt{f2}) locks the door, sends a message notification, and transmits out the door lock state to a remote server (i.e., http://support.com) (\circled{3}). However, the actual behavior of the app adds a piece of code that invokes a function (i.e., \texttt{leakinfo()}) sending a \emph{string} that contains ``Nobody is Home'' to a hard-coded phone number (\circled{4}). This string is highly private and informs an adversary that the house is empty. This information can be abused, for instance, to facilitate a burglar to break into the house. This example shows that a user does not have control over what an IoT app does with the sensitive data, who sees it, and what they do with it. Unfortunately, IoT development platforms do not provide users with sufficient information to make informed decisions about their privacy preferences in IoT environments. 

\vspace{3pt}\noindent\textbf{Limitation of Existing Privacy Tools.}
There exist systems to identify sensitive data-flows in IoT apps. For instance, \saint is a static system that uses taint analysis to identify sensitive data-flows in IoT apps~\cite{saint}. FlowFence, a dynamic system, uses quarantined modules to enforce data-flow policies on the use of sensitive data~\cite{fernandes2016flowfence}. However, these approaches are limited in precision and the number of privacy policies enforced. For instance, \saint and FlowFence have no way of knowing if an app leaks sensitive data through developer- or user-defined strings. For instance, the string ``Nobody is home'' leaked through \texttt{leakinfo()}). Meanwhile, our analysis of 540 IoT market apps showed that 64\% of apps potentially leak sensitive data through strings that do not include any tainted data, yet the string is sensitive. Lastly, static systems like \saint, iRuler~\cite{iruler}, and privacy tools for trigger-action platforms~\cite{ifttt} fail to detect sensitive data leaks from methods defined dynamically in IoT apps~\cite{groovy-dynamic}. Additionally, there are no approaches that allow users to examine their sharing and privacy preferences over individual IoT apps. For instance, a user may desire to share her energy usage data with a third party in a specific IoT app yet she wants to restrict sharing all other sensitive information with third-parties. This requires a personalized privacy setting for each app that gives control to the users over what to share.

In contrast to previous approaches, \daint analyzes data-flows at runtime; thus, the flow's content is analyzed to determine whether a data-flow constitutes a privacy concern or not. Such a runtime analysis capability overcomes the limitations of static analysis tools that fail to consider taint variables generated dynamically. Additionally, \daint provides additional user interfaces that allow users to configure their privacy settings for each app and informs the users about its findings. We provide a detailed comparison of \daint with other privacy tools for IoT and Android apps in Section~\ref{sec:relatedwork}. 

\vspace{3pt}\noindent\textbf{Threat Model and Assumptions.} We consider sensitive data leaks in IoT apps through malicious apps or unintentional developer mistakes. We consider sensitive information leaks in IoT apps via messaging and Internet connections, apps that transmit data to the recipients that are not authorized by users, or apps that transmit sensitive data without proper data protection mechanisms implemented. We do not consider safety and security violations in IoT apps~\cite{Berkay2018SoteriaUSENIXATC, Berkay2019IoTGuardNDSS}. Additionally, we do not track data-flows via push notifications or \textit{sink functions} that are authorized by OAuth (e.g., a user authorizes a third-party service through OAuth protocol to share the device states for data visualization).
%Leo: I cleaned up above.

%%%%%%%%%%%%%%%%%%%%%%%%%%%%%%%%%%%%%%%%%%%%%%%%%%%%%%%%%%%%%%%%%%%%%%%%%%%%%%%%%%%%%%%%%%%%%%%%%%%

\section{Approach Overview}
\label{sec:overview}

\begin{figure}[!t]
\centering{\includegraphics[width=0.4\textwidth]{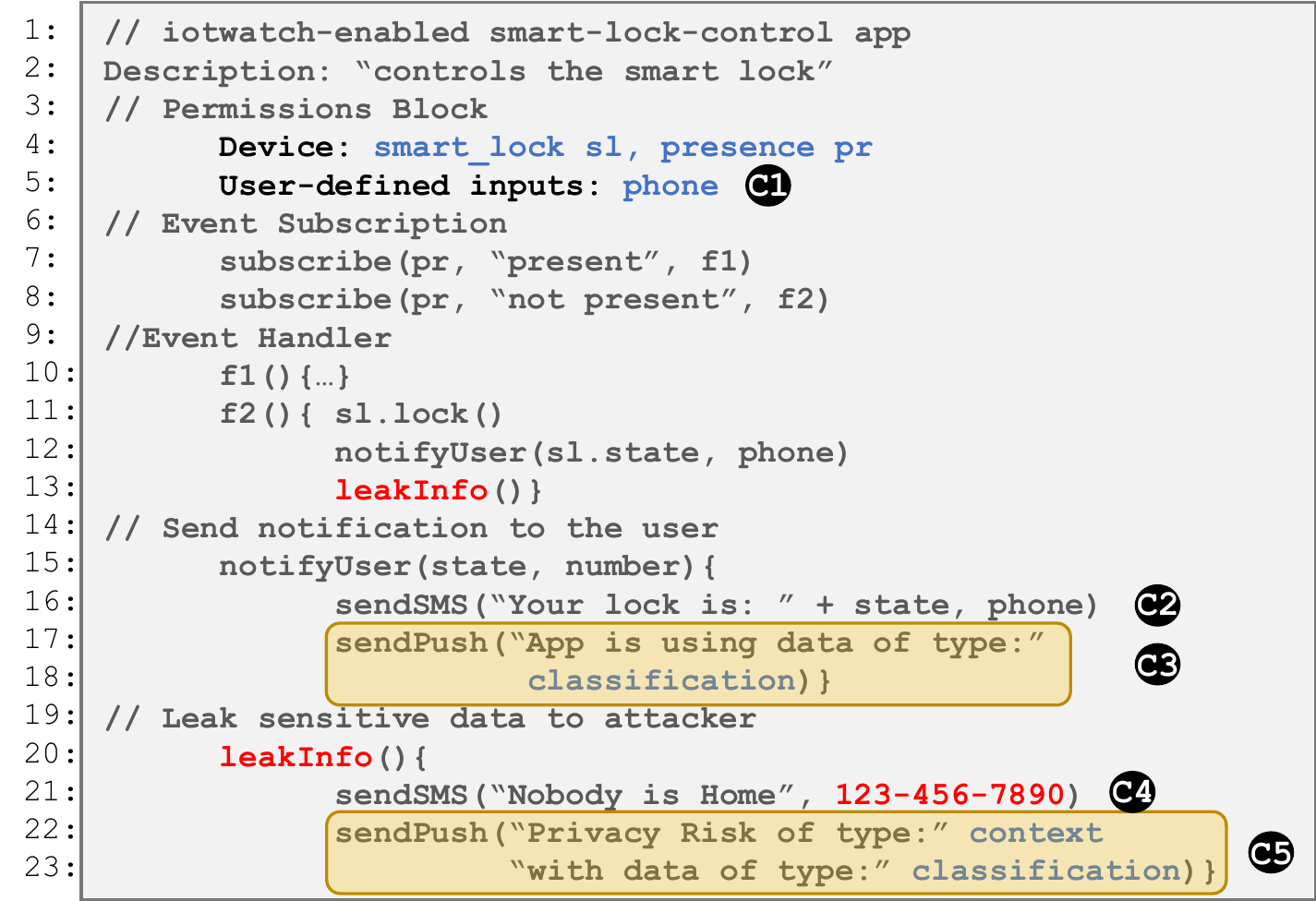}}
    \caption{An example of an IoT app instrumented by \daint to support the notification interface.}
    \vspace{-0.2in}
\label{fig:daintedAppCode}
\end{figure}

\daint, a runtime privacy analysis system, collects the information exchanged with external parties in IoT apps to uncover privacy risks. The collected information is used to classify sensitive data into easy-to-understand privacy labels. The findings are checked against users' privacy preferences and users are notified if there is a mismatch. \daint helps users define their privacy preferences and have better control over their sensitive data and maintain better privacy practices.

\begin{figure*}[t!]
\centering{\includegraphics[width=0.85\textwidth]{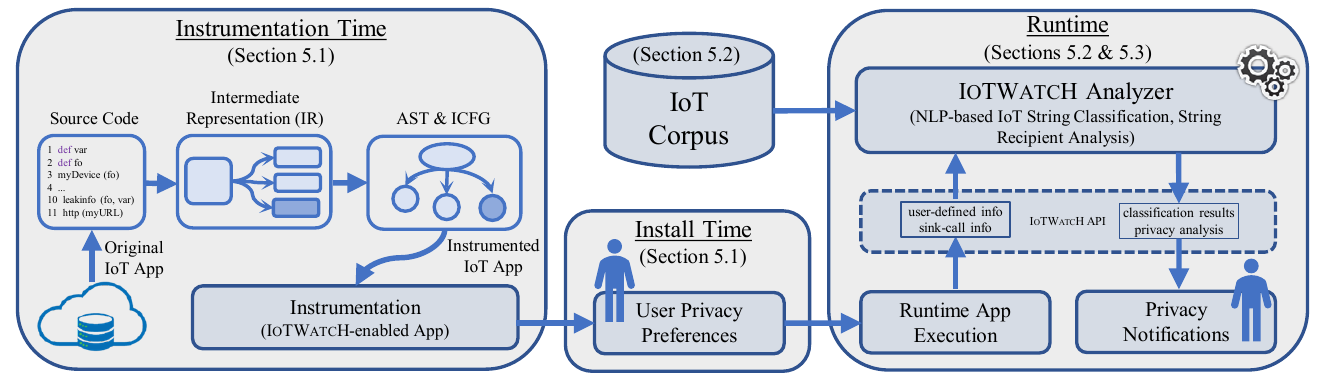}}
    \caption{Overview of \daint architecture. Three main stages are highlighted: first, IoT apps are modified at instrumentation time to enable \daint; second, the user selects their privacy preference at install time; finally, at runtime, \daint analyzes the IoT app data to uncover privacy risks and behaviors.}
    \vspace{-0.2in}
\label{fig:architecture}
\end{figure*}

\subsection{Understanding Leakage in IoT Apps}
We use the \texttt{smart-lock-control} app depicted in Figure~\ref{fig:originalApp} to illustrate the logical steps of \daint (Figure~\ref{fig:daintedAppCode}). The instrumentor first adds extra logic to the app's source code to implement a user interface (UI). The UI allows a user to select a set of privacy labels and the type of potential leakage mechanisms to be tracked by \daint. We provide users with four privacy labels: location, user-behavior, device-info, and date-time, and two potential leakage mechanisms: messaging and Internet, to define their privacy profile. Additionally, the code instrumentor adds extra logic to send app data to \daint server at runtime. The data includes user-defined input variables (i.e., phone) (\circled{\footnotesize{C1}}) and the content and recipients of the functions used to send data out of the app. The sample code in the \texttt{smart-lock-control} app includes two messaging functions (\circled{\footnotesize{C2}} and \circled{\footnotesize{C4}}). \daint instruments both the messaging functions to collect their content (i.e., ``Your lock is: '' + state, and ``Nobody is Home'') and their recipients (i.e., phone and 123-456-7890). Lastly, the instrumentor inserts extra code to enable the notification of \daint's results to users (\circled{\footnotesize{C3}} and \circled{\footnotesize{C5}}).

At install-time, a user configures two pieces of privacy settings for the app based on her privacy preferences: (1) the privacy labels that are of interest of the user and (2) the types of possible leakage methods to be tracked by \daint. These privacy settings define the user's privacy profile and are used to notify the user when her sensitive data is transmitted out of an app. The user then installs the instrumented app, which transmits its data to the \daint server once a specified data-flow is flagged. This information enables \daint to identify the type of sensitive information that the IoT app used and to analyze the information obtained from the app to uncover privacy data leaks and potential privacy behaviors. \daint implements a novel algorithm to verify whether the sensitive data is sent to recipients defined by the user via matching the user-defined inputs to the data recipients. For instance, for the first potential leakage (\circled{\footnotesize{C2}}), the data is sent to a user-defined phone number (\circled{\footnotesize{C1}}). For the second possible leakage (\circled{\footnotesize{C4}}), the data is sent to a hard-coded phone number, which might indicate a privacy violation to the user. Also, by analyzing the content of these potential leakages with NLP-based techniques, \daint verifies that the app uses sensitive information regarding the devices (i.e., ``Your lock is:'') and location (i.e., ``Nobody is Home''). Lastly, \daint informs the user of its findings and generates privacy awareness. For the first possible leakage, \daint sends a push notification with the privacy labels of the information included in the message (\circled{\footnotesize{C3}}). For the second potential leakage, \daint informs the privacy content of the message and also alerts the user regarding the potential privacy violations of the user's privacy (\circled{\footnotesize{C5}}). 

\subsection{Terminology Used}
\label{sec:terms}
We define some of the terms used to explain \daint's architecture. 

\vspace{0pt}\noindent\textbf{Sink-calls.} IoT programming platforms define specific APIs to send information out of the apps (i.e., sink-calls)~\cite{smartThings-documentation, openhab} as external data-flows. In this work, we focus on sink-calls of type messaging and Internet. Sink-call methods require two types of information: the \textit{recipient} and the \textit{content}. Specifically, the recipients define where the information contained in the sink-call is being sent to, and the content defines the message or data sent in the form of IoT strings. 

\vspace{0pt}\noindent\textbf{Privacy Labels.} We define four different privacy labels (i.e., date-time, device-info, location, and user-behavior) to classify IoT strings (i.e., sink-call content) in IoT apps that users can specify. These privacy labels were selected based on the findings and takeaways extracted from a privacy survey as noted in Section~\ref{sec:userstudy}. 

\vspace{0pt}\noindent\textbf{Privacy Profile.} We consider the collection of privacy labels and notification preferences defined by the \daint's user as a privacy profile. At install-time, the user selects the preferred type of privacy information and communication methods (e.g., messaging, Internet) to be tracked by \daint.  

\vspace{0pt}\noindent\textbf{Privacy Leakage.} We consider any data that is sent to an external recipient that is not authorized (i.e., defined by the user at install-time) or acknowledged (i.e., informed to the user via the app's description block) by the user of the IoT app as a leakage. 

\vspace{0pt}\noindent\textbf{Privacy Behavior.} We consider any sink-call that sends data to an authorized recipient, but that potentially puts the sensitive information at risk as a privacy behavior. For instance, Internet communications sending the information to legitimate servers in plain text.

%%%%%%%%%%%%%%%%%%%%%%%%%%%%%%%%%%%%%%%%%%%%%%%%%%%%%%%%%%%%%%%%%%%%%%%%%%%%%%%%%%%%

\section{\textsc{IoTWatcH}}
\label{sec:architecture}
Figure~\ref{fig:architecture} illustrates \daint's architecture which includes processes performed in three main phases: instrumentation time, install time, and runtime. At instrumentation time, the code instrumentor adds extra logic to the app's source code (1) to implement a privacy interface (Section~\ref{sec:selective}) where users specify their privacy preferences, and (2) to collect app information required for \daint analysis. At install time, the user defines the type of privacy information they desire to be tracked and notified about by \daint. Finally, at runtime, the app sends its information to \daint server, which classifies the collected data into privacy labels through NLP techniques (Section~\ref{sec:NLP}). In addition, \daint performs analysis on the data recipients  (Section~\ref{sec:contextAnalysis}) for potential privacy leakages. In the case of sink-calls of type messaging, it matches their recipients to user-defined inputs to check if the apps send sensitive information to unauthorized parties. In the case of Internet communications, it checks whether an app sends information to the remote servers using unencrypted HTTP calls. \daint's classification process is supported by a model implemented from a market IoT corpus. Lastly, \daint informs the user about its findings (Section~\ref{sec:response}). 

\begin{comment}
\begin{figure}[!t]
\centering{\includegraphics[width=0.4\textwidth]{Figures/Instrumenter-6.pdf}}
    \caption{Logical steps of \daintf's IoT app instrumentation.}
    %\vspace{-0.2in}
\label{fig:instrumenter}
\end{figure}
\end{comment}

\subsection{Code Instrumentor}
\label{sec:codeInstrumentor}
\daint's code instrumentor analyzes the source code of an original IoT app to build an intermediate representation (IR). The IR allows one to extract the sensor-computation-actuator paradigm of IoT apps~\cite{saint, Berkay2018SoteriaUSENIXATC}. The use of an IR enables the design of generic solutions that can be implemented for different IoT programming platforms (e.g., Smasung SmartThings and OpenHAB)~\cite{saint}. Then, the instrumentor extracts the Abstract Syntax Tree (AST) of the app and implements custom \textit{node visitors} to build the Inter-procedural Control Graph (ICFG). The ICFG is used to flag user-defined inputs in the permission block of the app, and the recipients and content of the sink-call functions (i.e., messaging and Internet). Then, the instrumentor adds extra code to collect and transmits this data to the \daint's server, and to implement push notifications that informs the user about \daint's findings in real-time. The code instrumentor groups collected data into two different categories: (1) app information and (2) sink-call information. We detail each of them as follows:

\vspace{3pt}\noindent\textbf{App Information.} \daint's code instrumentor visits the permission block (Figure \ref{fig:daintedAppCode}) in IoT apps and extracts information defined by the user. Specifically, it collects user-defined inputs required to implement the sink-calls (\eg phone numbers to receive notifications from the app). As we detail in Section~\ref{sec:analyzer}, this data is matched with the information directly extracted from sink-calls (i.e., recipients) to detect, for instance, data sent to unauthorized recipients (i.e., not defined by the user), which may lead to potential privacy issues for the user.

\vspace{3pt}\noindent\textbf{Sink-Call Information.} It includes the content and recipients of messaging and Internet communications. For instance, in the app's source code depicted in Figure \ref{fig:daintedAppCode}, \daint extracts the content ``Your lock is: + \textit{state}'' from the first messaging function in Line 16, and the content ``Nobody is Home'' from the second messaging function in Line 21. As per the recipients, it extracts the recipient's value contained in the variable $phone$ from the first messaging method (Line 16) and the hard-coded phone number ``123-456-7890'' from the second messaging function (Line 21). \daint uses the content of the sink-calls to make the user aware of the type of sensitive information that IoT apps handle. We use an NLP-based model to analyze the content of external data-flows and classify them into four privacy labels that are easy to understand by the user. Also, it uses the recipient information in \daint analyzer to uncover sensitive data leaks. Our tool matches the recipient information extracted from messaging and Internet communications with both the user-defined and user-acknowledged information at install-time. The user-defined information is entered by the user (Line 5) and the user-acknowledged information is informed by the developer via the app's description block (Line 2) and approved by the user. In cases where the sink-call (i.e., messaging or Internet) is executed using unauthorized recipients, the flow is flagged as \textit{leak}, and the user is informed. With this analysis, \daint creates awareness of sensitive information being disclosed to unauthorized or malicious recipients. Besides, for Internet communications, we verify that the recipient supports data encryption. With this, our tool guarantees that the sensitive information is protected from potential eavesdroppers. 

\begin{figure}[t]
\centering
    \subfigure[Original IoT app]{\includegraphics[width=0.19\textwidth]{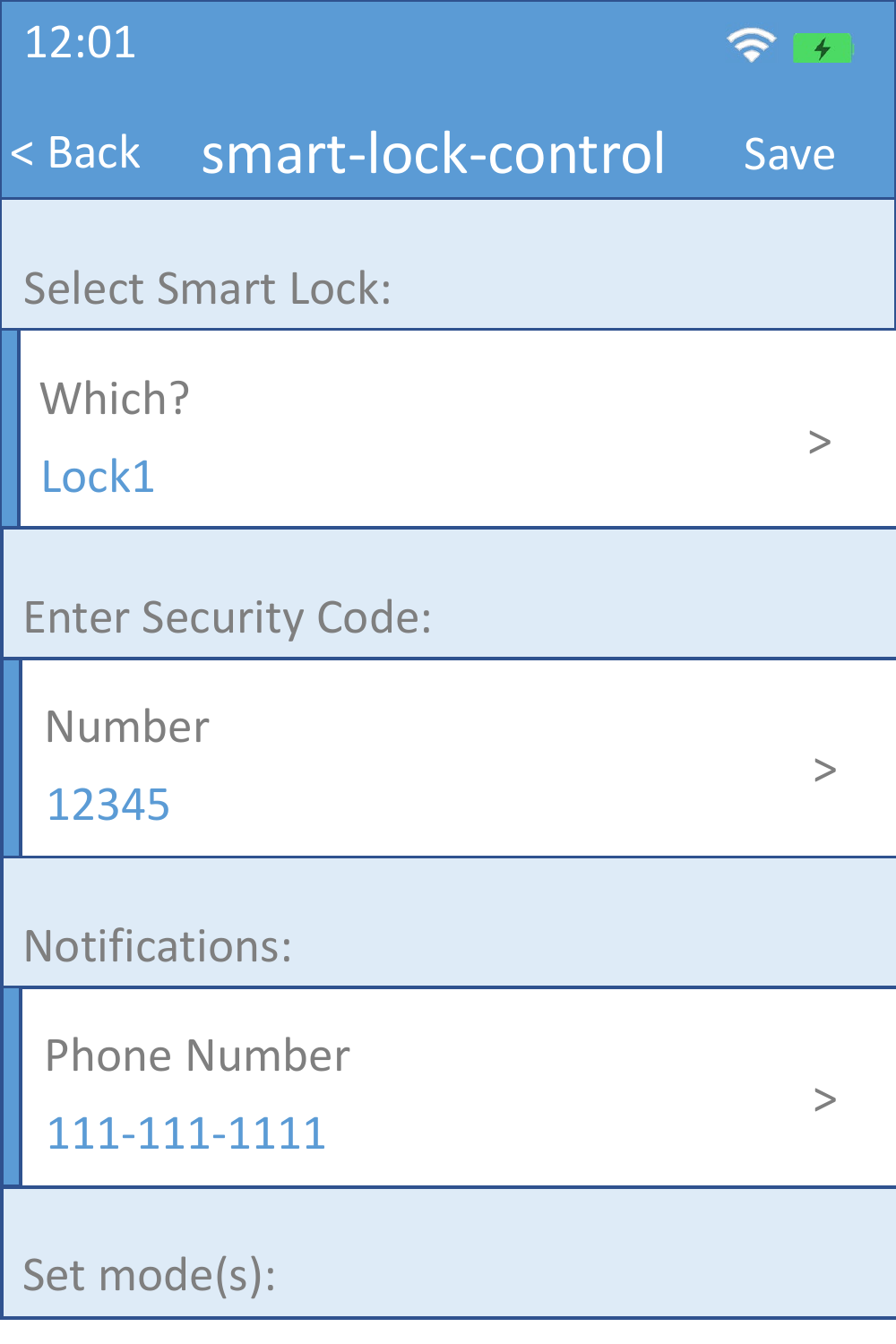}\label{fig:OrigApp}}
    \subfigure[Instrumented IoT app ]{\includegraphics[width=0.19\textwidth]{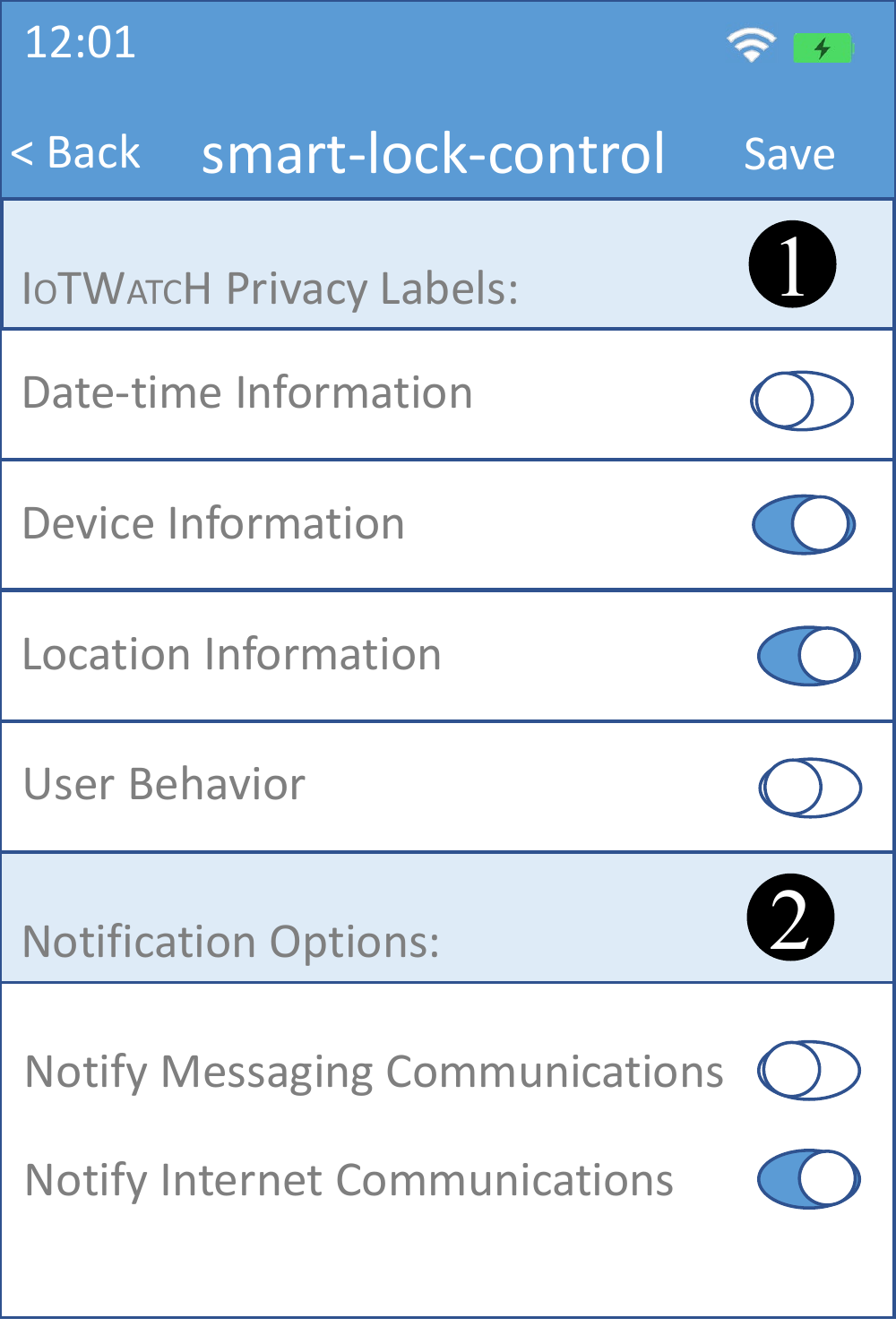}\label{fig:daintedApp}}
    \vspace{-0.1in}
    \caption{(a) Install-time interface of an IoT app and (b) Instrumented IoT app interface: \daintf interface enables users (1) to select privacy labels and (2) to identify unauthorized recipients when a sensitive information is leaked.}
    \vspace{-0.2in}
\end{figure}

\subsubsection{Selective Instrumentation}
\label{sec:selective}
In addition to collecting app information, \daint performs a selective code instrumentation to support on-demand privacy analysis/notifications and to facilitate the analysis of encrypted IoT strings. 

\vspace{1pt}\noindent\textbf{Privacy User Interface.} The instrumentor adds additional code to implement a UI and create a privacy profile of the user. Figure~\ref{fig:OrigApp} shows the original user interface of an IoT app presented to the user during install-time, and  Figure~\ref{fig:daintedApp} illustrates the selective code instrumentation options of the \daint-enabled app. \daint's instrumentor does not impact the UI experience of the IoT app at runtime, but offers new privacy features at install-time not available in the original app. The instrumented app offers the user the possibility to create a privacy profile and receive notifications regarding specific privacy labels that are of interest to the user (\circled{1}). Also, it allows for selecting which privacy concerns (e.g., option to notify leaks from messaging or Internet communications) must be analyzed and informed by \daint (\circled{2}). Such a design supports the expectations of the IoT app users with (1) configurable privacy preferences, (2) on-demand privacy controls, and (3) timely privacy notifications (Table~\ref{tab:features}) as were summarized in Section~\ref{sec:userstudy}. Finally, since we target open-source IoT platforms, the use of selective instrumentation does not change or impact the original user interface of the app that may be considered as intellectual property. For closed-source platforms, we envision developers using the features offered in \daint to evaluate and improve the protection of sensitive information and the privacy of users.

\begin{figure}[!t]
\centering{\includegraphics[width=0.4\textwidth]{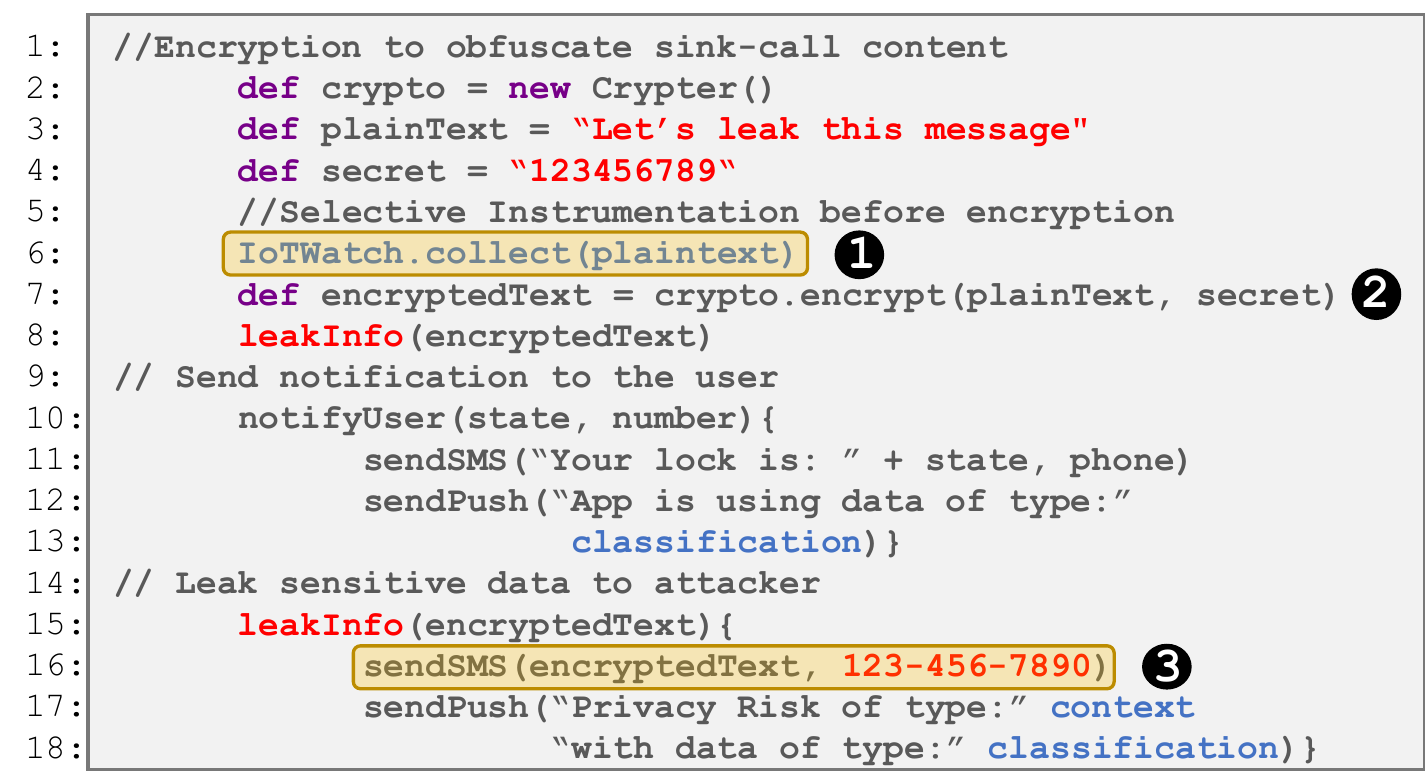}}
    \caption{Sample IoT app that encrypts the sensitive data to be leaked in an attempt to bypass the NLP analysis of \daint. The selective instrumentation capabilities of \daint permits the analysis of the data before it is encrypted.}
    \vspace{-0.2in}
\label{fig:encryption}
\end{figure}

\vspace{1pt}\noindent\textbf{\daint vs. Encryption.} The use of encryption to hide the content of a sink-call may limit the effectiveness of NLP techniques. For instance, NLP models created from plain-text data would fail to classify encrypted strings. However, selective instrumentation may facilitate the analysis of IoT apps that codify or encrypt the data that is sent out to hide their intent. In general, IoT programming platform only permits limited (i.e., white-listed) number of libraries to implement the app's execution~\cite{smartthingsJava}. In popular IoT platforms like Samsung SmartThings, this group of white-listed APIs does not include a single class that implements encryption. Also, as a general rule, IoT platforms reject obfuscated apps when developers submit them for approval during the vetting process. However, there still several reasons to allow encryption-handling in \daint. First, malicious apps may skip the vetting process of IoT programming platforms and while still be in the app market. Second, encryption handling allows for a privacy tool that is not only exclusive to specific IoT platforms, but that can be generalized as broader solution. As explained before, the use of a limited (i.e., white-listed) number of libraries to code the IoT apps facilitates the detection of encryption functions by \daint. Figure~\ref{fig:encryption} shows a sample IoT app that implements encryption to hide its intentions of leaking a sensitive string via messaging. The method \texttt{crypto.encrypt} is known to be approved to implement encryption. \daint's selective instrumentation extracts the call graph of the app and detects the presence of a function implementing encryption (\circled{2}) on a tained variable \texttt{plainText}. The result of the encryption is stored in a new taint variable \texttt{encryptedText}, which is later leaked via messaging (\circled{3}). To solve this challenge, the instrumentor tracks the flow's path of the encrypted variable and inserts extra code to collect its content before it is encrypted (\circled{1}).  

\begin{comment}
\begin{figure}[!t]
\centering{\includegraphics[width=0.3\textwidth]{Figures/CGF.pdf}}
    \caption{Example call graph extracted and selectively instrumented by \daint. Our tool tracks the variable containing the sensitive data and adds extra code to collect its value before it is encrypted.}
    \vspace{-0.2in}
\label{fig:cgf}
\end{figure}
\end{comment}

During our analysis, out of 540 current market and malicious IoT apps, we did not find a single case of an app encrypting the content while using messaging. However, we found several cases of encryption in Internet communications. \daint's instrumentor bypasses encryption on the content of the Internet calls by analyzing the entire data-flow path within the apps (i.e., from the data source to the sink functions), and by collecting the information related to the sink-call content before it is encrypted or codified. In the specific case of encrypted Internet communications, the instrumentor flags the call graph nodes containing the \texttt{https} call methods and adds additional code to collect its content before the encryption call is executed.

\subsection{\textsc{IoTWatcH} Analyzer}
\label{sec:analyzer}
\daint uses NLP techniques to analyze and classify an app's sink-call contents to four privacy labels. Additionally, it flags messaging and Internet communications that disclose sensitive information to unauthorized recipients. Finally, it uncovers privacy concerns from apps that do not protect the sensitive information from passive observers. Figure \ref{fig:architecture} illustrates the logical steps of the \daint's analyzer at runtime, that includes: (1) classification of sink-call content through NLP techniques, (2) privacy analysis of recipients of sensitive information, and (3) building a user privacy notification interface to inform the user about \daint's findings. 

\subsubsection{Classification of Sink-Call Content}
\label{sec:NLP}
The content of messaging and Internet communications in IoT apps may include sensitive information from taint variables or might be just string messages with privacy implications to the user. As noted earlier, the sink-call content classification of \daint takes the content of messaging and Internet communications as input and assigns it to privacy labels returned as the output. For instance, if the sink-call contains the string message ``the door is unlocked'', the classifier performs semantic analysis on the string and classifies it as ``Device-info''. The conversion of IoT strings into privacy labels helps the user to understand how IoT apps use sensitive information so they can make informed privacy decisions. Below, we present how to construct a training set for classification from a corpus of IoT apps. %First, we assign privacy labels to each sample data consisting of a string extracted from messaging and Internet communications in real-time. Then, we train the model to classify unknown samples to privacy labels. 

\vspace{1pt}\noindent\textbf{Constructing Privacy Labels.} \daint classifies IoT app sink-call contents into a set of privacy labels. The use of privacy labels provide three main advantages. First, they allow users to understand what type of data the apps leak. For instance, a string that contains ``The mode has been changed to Home'' can be presented to users as leaking location. Second, the privacy labels permit users to have control over their privacy preferences. For instance, a user may desire to allow an app to transmit energy usage of a thermostat to a remote server, yet she restricts other types of privacy-related information. Third, the privacy labels enable an on-demand notification system that guarantees an improved user experience with less disruptive and more intuitive notifications, and minimal runtime delay added to the app's execution time. 

\begin{comment}
\begin{figure}[!t]
\centering{\includegraphics[width=0.45\textwidth]{Figures/Architecture-6.pdf}}
    \caption{The components of the \daint analyzer.}
    \vspace{-0.2in}
\label{fig:analytics}
\end{figure}
\end{comment}

We define four privacy labels through the analysis of the semantics of strings extracted from messaging and Internet communications in IoT apps (although adding more privacy labels later is a straightforward task). The privacy labels are based on user feedback that we acquired via a privacy survey (Section~\ref{sec:userstudy}). Also, to adequately protect the user's privacy, some strings may require the use of more than one privacy label to guarantee completeness on the classification. For instance, the message string ``The door will remain open for another 5 minutes'' supports multi-labeling of types Device-info and Date-time. Similarly, ``Garage door is not opening since the car was not present at Home, less than 15 sec ago'', would be labeled as Device-info (the door is not opening so it is still closed), Location (the car was not present at Home), and Date-time (15 sec ago). Some examples of IoT sink-call contents assigned to the various privacy labels are presented in Table~\ref{tab:flows}. In general, the use of multiple privacy labels to a single text considers more complex semantics structures of string and reveals more privacy-sensitive information contained in the leaked message. As we explain later in this section, we successfully use NLP to achieve this goal. We present the four privacy categories used to classify IoT strings:

\begin{itemize}
\itemsep0em 
\item \vspace{1pt}\noindent\textit{Date-time:} defines an app text that contain time or date information. For instance, a messaging call that sends the string ``door is unlocked at 5:00 pm" contains time information that specifies the time that the door would be unlocked. We found that many IoT apps contain date-time information for reporting the state of a device at any given time.

\item \vspace{1pt}\noindent\textit{Device-info:} defines the text that contain the states of devices and also device information (i.e., device type, model, manufacturer). For instance, an Internet call that contains ``energy usage''  transmits out the power state of a thermostat.  We assign all strings that contain information from devices with a device-info label to inform the users about potential privacy violations.

\item \vspace{1pt}\noindent\textit{Location:} defines the text that reveal physical and geo-location location of users and devices. For instance, the string ``kids have arrived home'' contains physical location information, and ``the patio door is unlocked'' contains the geo-location of a house.

\item \vspace{1pt}\noindent\textit{User-behavior:} defines the text that provides information about user preferences. We found that many IoT apps leak strings in messaging calls about app configuration and user activities. For instance, an app includes a string, ``the user mode changed to vacation from home''. %We label these strings with user-behavior as it provides information about user preferences.
\end{itemize}

\begin{table}[!t]
\centering
\resizebox{\columnwidth}{!}{
{\scriptsize{
\begin{tabular}{|lc|}
\hline
\textbf{\begin{tabular}[l]{@{}l@{}}App Sink-Call Content\end{tabular}} &  
\textbf{\begin{tabular}[c]{@{}c@{}}Assigned Privacy Labels\end{tabular}} \\ 
\hline \hline
\rowcolor{light-gray}
\begin{tabular}[l]{@{}l@{}}Thermostat is turned on.\end{tabular} & device-info \\   
%\rowcolor{light-gray}
\begin{tabular}[l]{@{}l@{}}The door will remain open for another\\ 5 minutes.\end{tabular} & \begin{tabular}[c]{@{}c@{}}device-info,\\ date-time \end{tabular} \\   
\rowcolor{light-gray}
\begin{tabular}[l]{@{}l@{}}Door is not opening since car was not\\ present at Home, less than 15 sec ago.\end{tabular} &\begin{tabular}[c]{@{}c@{}} device-info,\\ location, \\date-time  \end{tabular}\\ 
\begin{tabular}[l]{@{}l@{}}Sleep time set for you as requested.
\end{tabular} & user-behavior \\ 
\hline
\end{tabular}
}}}
\vspace{-0.1in}
\caption{Examples of leaked strings extracted from IoT apps and their assigned privacy labels. Observe that \daint is capable of assigning multiple privacy labels to specific strings with more complex semantics.}
\vspace{-0.2in}
\label{tab:flows}
\end{table}

\vspace{1pt}\noindent\textbf{IoT App Corpus.}
%\label{sec:corpus}
Our study of sink-call contents (i.e., IoT strings) from 540 real IoT apps showed that they pose a few unique characteristics compared to data-flows from other domains. First, the size of the texts extracted is usually three to four shingles on average, yet it contains highly private data (e.g., ``the door is unlocked''). Second, their linguistic structure is minimal regarding semantics (e.g., ``mode changed to away'') compared to other short documents extracted from popular general-knowledge corpora~\cite{corpora, google, wikipedia}. Third, their meanings usually are closely attached to the app's context (e.g., ``if the user is not-present, turn off the light''). Based on these facts, we designed an NLP model that can be effective for classifying semantic-deficient, but information-rich texts. To understand whether leaked strings includes sensitive information and assign them a privacy label, we first implemented a classification model using publicly available data corpora~\cite{corpora, google, wikipedia}; however, due to the specific characteristics of IoT texts, we obtained very low classification accuracy (more details in Appendix~\ref{sec:initialmodel}). To improve these initial results, we constructed an \textit{IoT-specific} corpus for classification purposes that successfully considers and solves the challenges above. To do so, we first collected the content of messaging and Internet calls from current IoT market apps. We pre-processed the resulting dataset by filtering out punctuation and stop words. Further, we manually labeled the IoT strings to the four privacy labels. Here, we applied multi-labeling to contents that contained information related to more than one privacy label. We detail constructing the IoT corpus in Section~\ref{sec:implementation}.

\vspace{1pt}\noindent\textbf{NLP Model Construction.} We implemented an NLP model that uses a specific IoT corpus for training purposes. Then, we used the model to classify unknown sink-call contents to privacy labels. To train the model, we used a supervised learning approach that required labeled data as input. We found that the supervised approach yields better results than other approaches such as keyword-search. This is because often, the IoT string does not include enough information to assign the labels through simple keyword-search-based analysis. For instance, keyword-search would fail to identify the user-behavior in a messaging communication that leaks a string ``the kids left home''. Moreover, we used \textit{skip-gram} and \textit{doc2vec}~\cite{word2vecbook} to represent every IoT corpus text into a multi-dimensional vector. We use doc2vec over other known approaches like bag-of-words (BoW)~\cite{bagofword} as it considers the entire structure of the text to perform syntax analysis (i.e., multi-word expression analysis)~\cite{doc2vec}. We then performed topic classification of the IoT strings using an automated machine learning approach~\cite{automl}. Note that the analysis of NLP results from different classification algorithms is out of the scope of this work. In fact, our automatic machine learning approach selects the best classification algorithm depending on the structure of text; thus, the best accuracy value is always guaranteed in \daint. In Section \ref{sec:implementation}, we provide details of the NLP implementation, including the IoT corpus used to train the model.

\subsubsection{Sink-Call Recipient Analysis}
\label{sec:contextAnalysis}
\daint performs sink-call recipient analysis via messaging by matching recipients of the data with information defined by the user at install-time. For instance, in Figure~\ref{fig:originalApp}, the user defines an authorized recipient in the variable $phone$. \daint extracts this information and correlates it with the app data extracted at runtime. In this specific example, two outcomes are possible. In Line 15, a messaging function is executed using $phone$ as the recipient. Since \daint recognizes that the user previously authorized this recipient, the flow does not represent a leakage. However, a second messaging function is executed in Line 19. In this case, \daint sees discrepancies between the recipient used (i.e., 123-456-7890) and the one that was defined by the user, and flags it as a leakage. Also, the analyzer checks the capabilities of the recipients of Internet communications to support encryption and adequately protects the sensitive information from, for instance, eavesdroppers. Back to our example, in Figure \ref{fig:originalApp}, an Internet communication is executed in Line 16. \daint extracts the recipient used (i.e., URL) and observes that the remote server is being accessed via Hyper Text Transfer Protocol (HTTP), so the information is sent out of the app in plain text. This constitutes a privacy behavior that could expose the content of the Internet message to passive observers (i.e., eavesdroppers). In this case, \daint flags this flow as a potential privacy concern for the user. 

\begin{figure}[t]
\centering
    \subfigure[User configuration]{\includegraphics[width=0.19\textwidth]{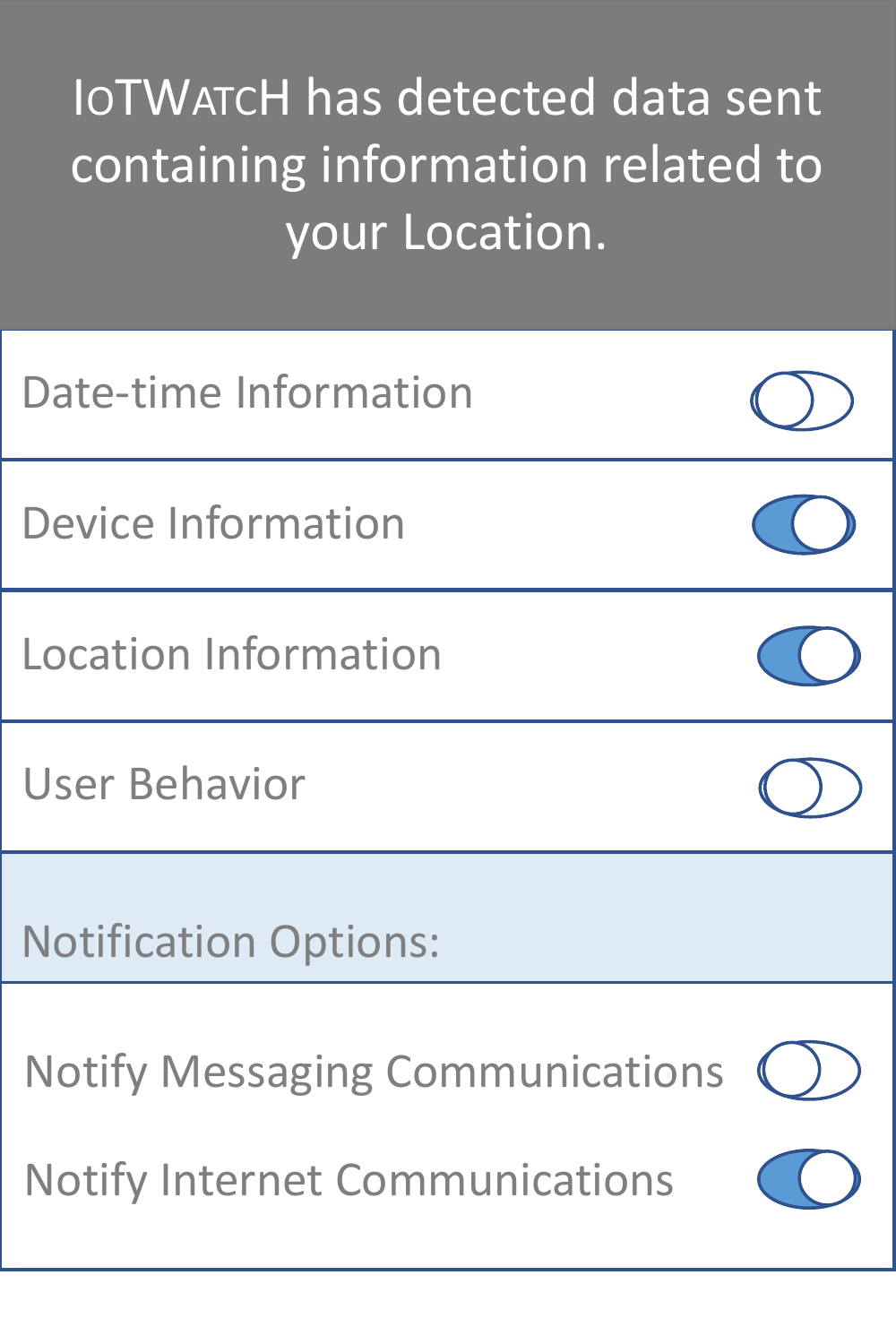}\label{fig:Classresult}}
    \subfigure[Notification to users]{\includegraphics[width=0.19\textwidth]{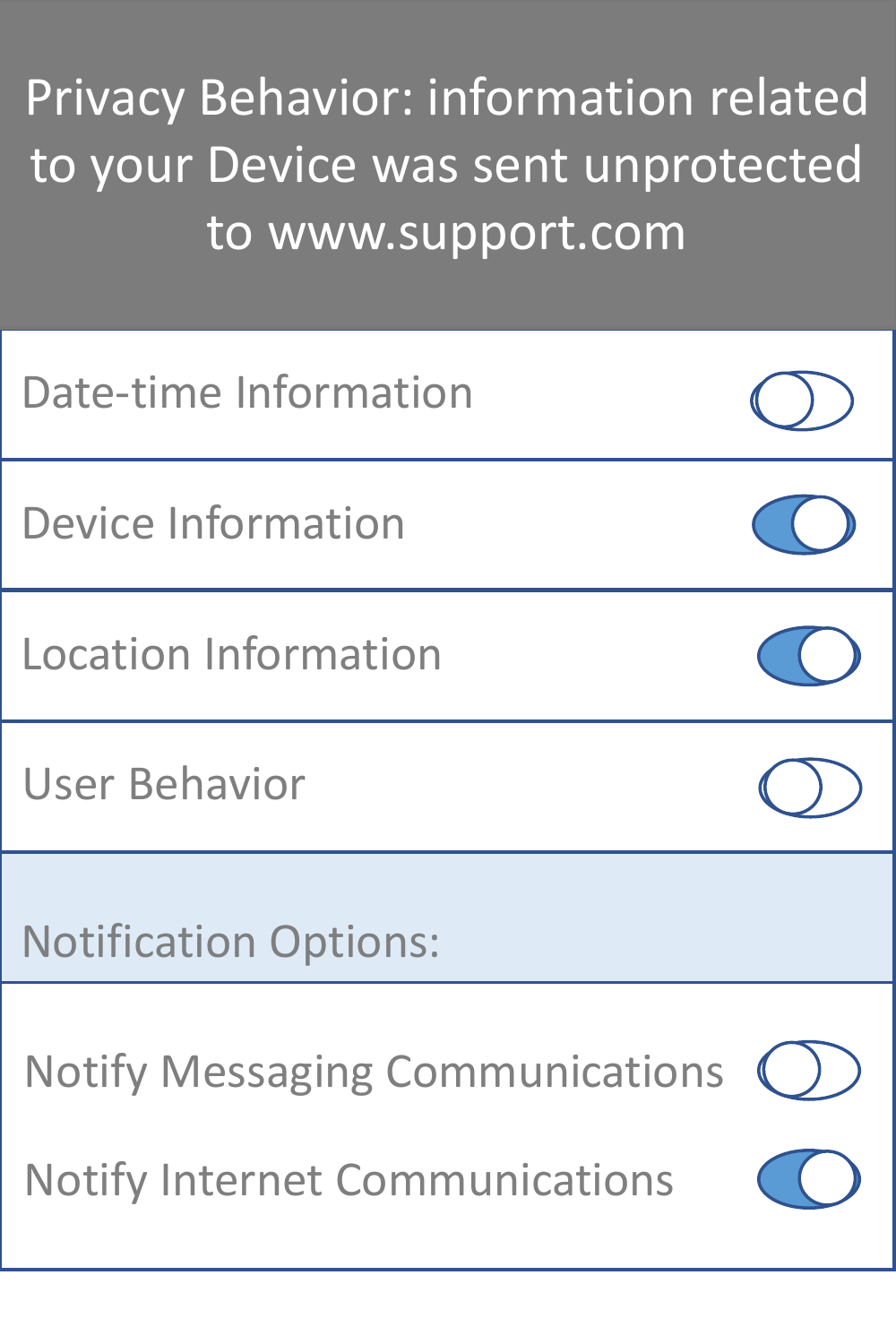}\label{fig:INTresult}}
    \vspace{-0.1in}
    \caption{\daint's findings are informed to the users through push notifications. The findings include (1) the privacy labels assigned to the sink-call content, and (2) the potential privacy concerns associated with the IoT communication.}
    \vspace{-0.2in}
\label{fig:results}
\end{figure}

\subsection{Response to App Data Leaks}
\label{sec:response}
Our privacy survey (Section~\ref{sec:userstudy}) shows that users of IoT apps desire on-demand privacy analysis and notifications. Based on this feedback, \daint implements two different privacy notification options (Figure~\ref{fig:results}). First, it allows a user to select specific privacy labels (one, multiple, or all) to create a privacy profile and receive notifications. For instance, if the user is only concerned about the use of data related to location information, she may select the location label so that \daint informs if a sink-call content contains information related to her and the devices' location. In this case, the user would not receive notifications regarding the use of other types of sensitive information within the IoT app. Second, \daint allows the user to decide if she desires to be notified regarding messaging, Internet calls, or both, whenever they potentially leak data to unauthorized parties. The approach of implementing a selective user-specific notification system provides flexible privacy options to the user, lowers the latency overhead of \daint by reducing the number of processed and classified IoT strings, and enhances the user experience by reducing the number of notifications at runtime. Finally, even though its flexible and very configurable nature, \daint enables all the privacy labels and notifications options by default. On the one hand, technically-enthusiastic users that fully understand the privacy risks of IoT apps may create their own privacy profiles by disabling install-time options in \daint, for a better user experience. On the other hand, users that are not aware of the privacy implications of sensitive information being leaked to third parties via IoT apps, or that do not completely understand the privacy labels in \daint, may rely on the default options.

\subsection{\textsc{IoTWatcH} API} \label{sec:api}
The analyzer collects app data to uncover data leaks and privacy behaviors in IoT apps. We implemented a REST API to enable effective data exchange and communication between the instrumented IoT app and \daint's analyzer running in the cloud (Figure \ref{fig:architecture}). From the app to the server, the API constructs a JSON object with the user-defined recipients and the sink-call information. From the server to the app, another JSON is transmitted including the analysis results and the notifications to the user (more details in Appendix~\ref{sec:apiappendix}).

%%%%%%%%%%%%%%%%%%%%%%%%%%%%%%%%%%%%%%%%%%%%%%%%%%%%%%%%%%%%%%%%%%%%%%%%%%%%%%%%%%%%%%%%%%%%%%%%%%%%%%

%\berkay{below is cleaned.}
\section{Implementation Details}
\label{sec:implementation}
We implemented \daint for IoT applications developed for Samsung \smarthings, which is the IoT platform that has the highest share of devices and applications in the current IoT market~\cite{smartThings-devices, smartThings-review}. Samsung \smarthings apps are developed in Groovy, a dynamic programming language that supports static compilation. Static compilation permits for all methods and classes in the apps to be annotated at compile time, which makes this information fully available to the instrumentation portion of \daint. 

\vspace{1pt}\noindent\textbf{Code Instrumentation.}
\daint traverses on the Abstract Syntax Tree (AST) of the IoT app's IR through the \texttt{ASTTransformation} class and builds an app's Intra-procedural Control Flow Graph (ICFG)~\cite{saint}. \daint involves around 1700 lines of code written in Groovy to analyze the app source code, construct the IR, generate the ICFG, and perform the code instrumentation. We implemented \daint's instrumentor as a web application using Groovy programming language (more details in Appendix~\ref{sec:web}). We made the instrumentor available online for the community.

\vspace{1pt}\noindent\textbf{Collection of IoT Corpus.} To avoid overlapping between the data used for training and evaluation of \daint, out of 540 apps, we selected 380 current market apps crawled from \smarthings repositories~\cite{Official, Community} to build the IoT corpus. The app population included apps from 6 different categories: \textit{Convenience}, \textit{Smart Home Automation}, \textit{Entertainment}, \textit{Personal Care}, \textit{Security \& Safety}, and \textit{Smart Transportation}. From the selected apps for training, we extracted a total of 2014 different IoT strings. We then labeled these strings according to the four privacy labels. Specifically, 46.8\% of the strings contained information related to the IoT devices, while 20.8\% contained relevant information related to Date-time. The remaining 19.2\% and 13.2\% of the IoT strings shared information related to location and user-behavior, respectively. Additionally, we allowed up to 75\% of inter-labeling assignment to the strings, meaning, up to three different privacy labels can be assigned to a single string.  
%Figure \ref{fig:corpus} depicts statistical details of the privacy label distribution used to build the IoT corpus. 
In total, we applied multi-labeling to 72\% of the privacy strings in the IoT corpus. Finally, we used 75\% of the total corpus to train the classifier. Then, we verified the obtained model with the remaining 25\% of the data. Initial testing results on the NLP model showed an average precision of 94.3\% and recall of 89.6\%.

\vspace{3pt}\noindent\textbf{Classification of IoT Privacy Strings.} We use Automatic Machine Learning (Auto-ML) tools offered by Google App Engine~\cite{automl} to perform privacy classification of IoT strings. Among its benefits, modern auto-ML approaches perform neural architecture search, enable hyper-parameter optimization, and utilize advanced model architectures to classify contents to privacy labels with high accuracy. More particularly, we implemented a custom multi-class multi-label model using the Natural Language API offered by Google (Google-NL)~\cite{googleautoml}. Google-NL offers a suite of ML algorithms that automatically optimize the algorithm parameters based on the specific algorithm utilized and the characteristics of the dataset to guarantee the highest accuracy. Our initial analysis of 2014 IoT strings showed remarkable semantic similarities among them; thus, the use of labeled data reduces the training time considerably. Finally, we noticed that only two (0.5\%) IoT strings in the corpus were in an idiom different from English (Spanish in both cases). Thus, we implemented our NLP solution to classify strings defined in English.

%%%%%%%%%%%%%%%%%%%%%%%%%%%%%%%%%%%%%%%%%%%%%%%%%%%%%%%%%%%%%%%%%%%%%%%%%%%%%%%%%%%%%%%%%%%%%%%%%%%%%

\begin{figure*}[!t]
\centering
    \subfigure[Accuracy]{\includegraphics[width=0.19\textwidth]{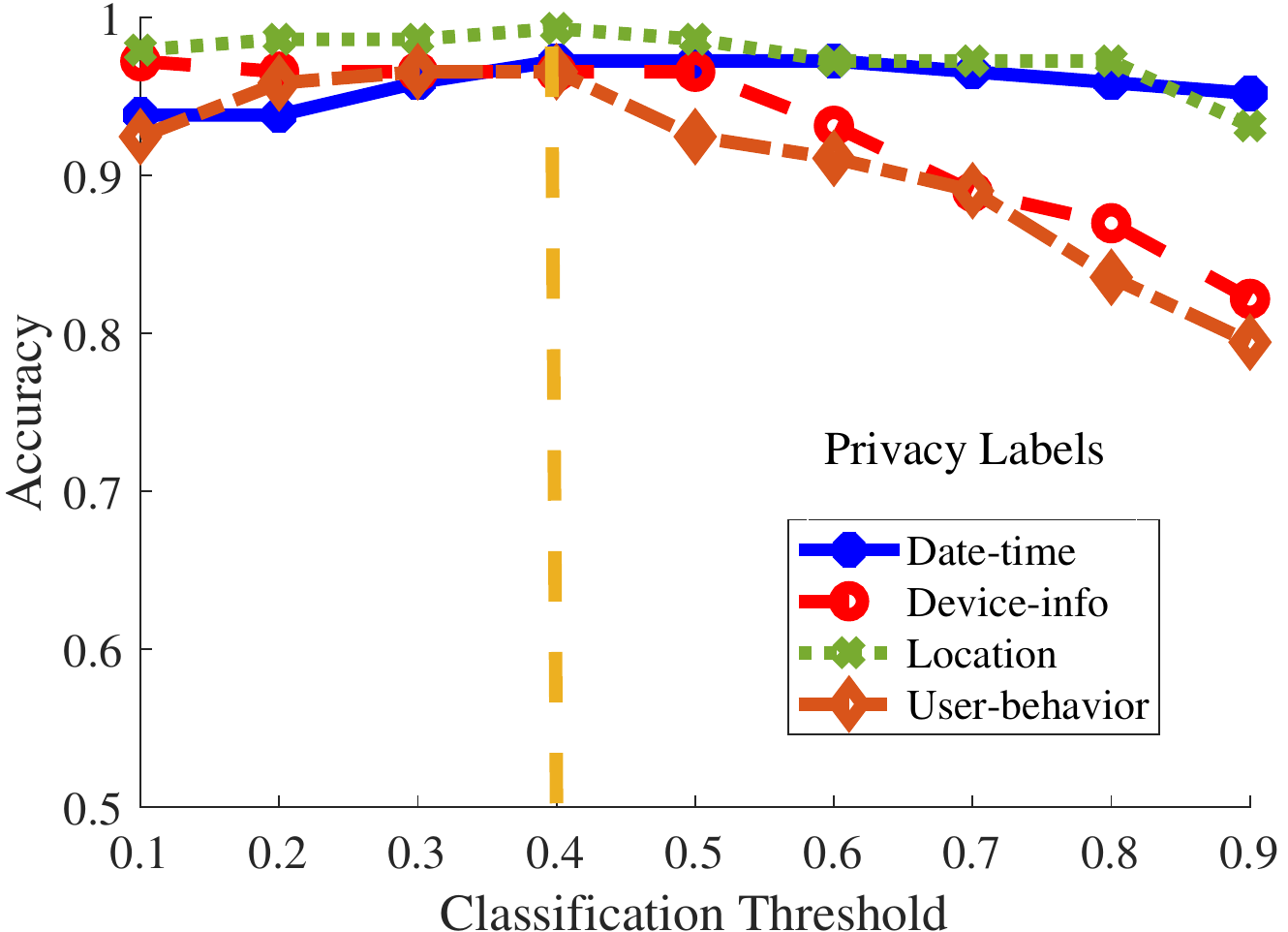}\label{fig:acc}}
    \subfigure[Recall]{\includegraphics[width=0.19\textwidth]{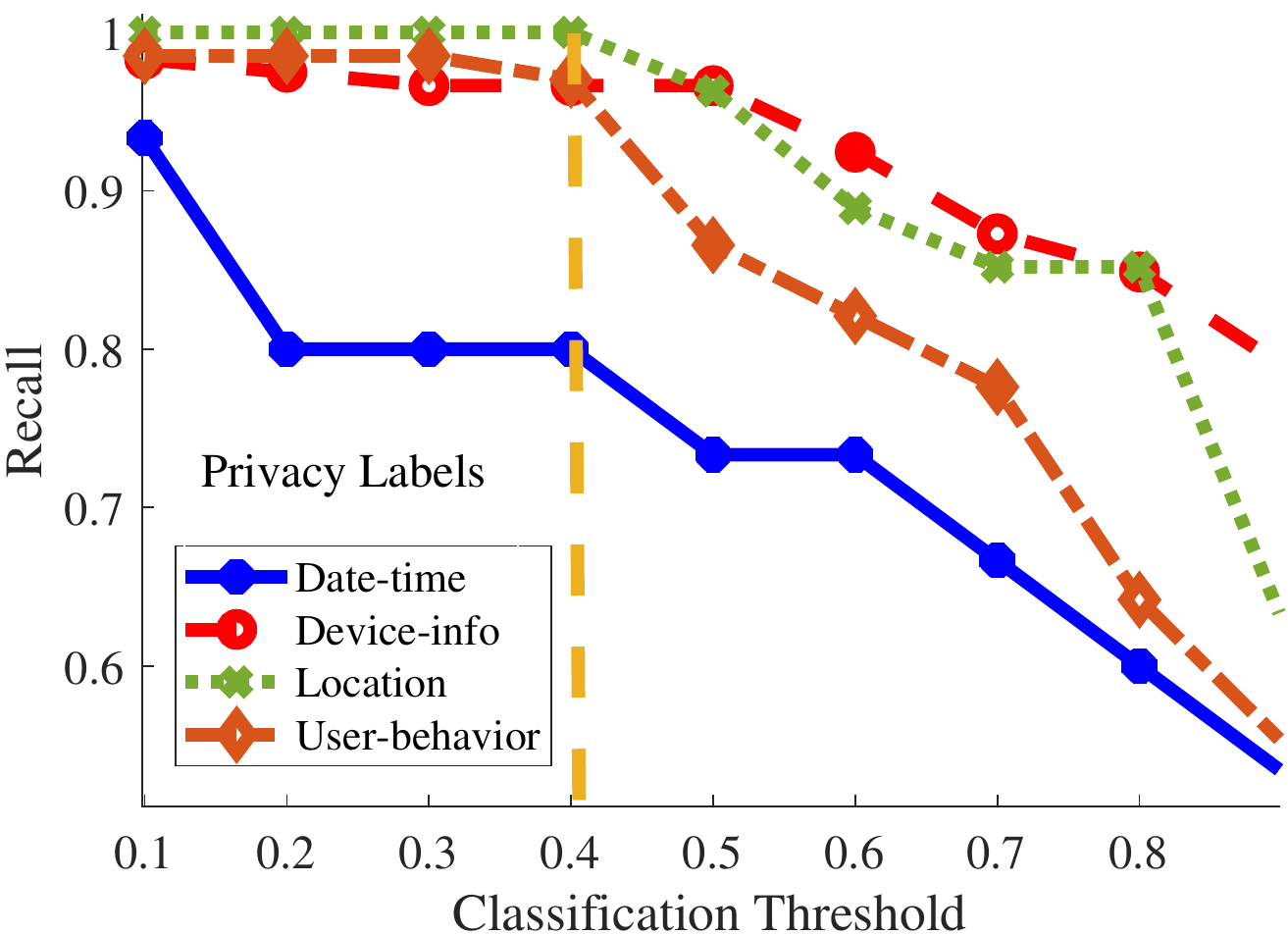}\label{fig:recall}}
    \subfigure[Precision]{\includegraphics[width=0.19\textwidth]{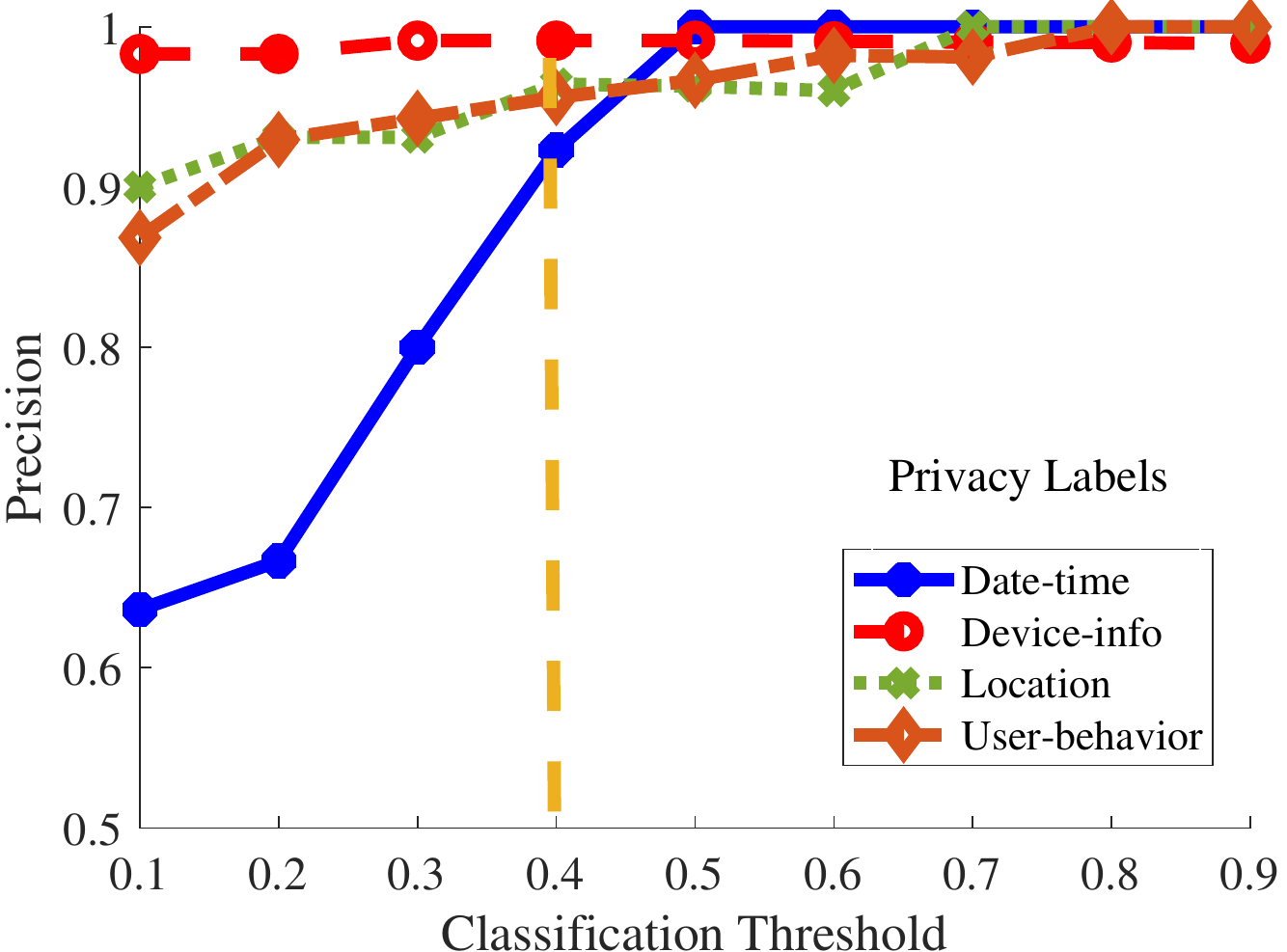}\label{fig:precision}}
    \subfigure[Specificity]{\includegraphics[width=0.19\textwidth]{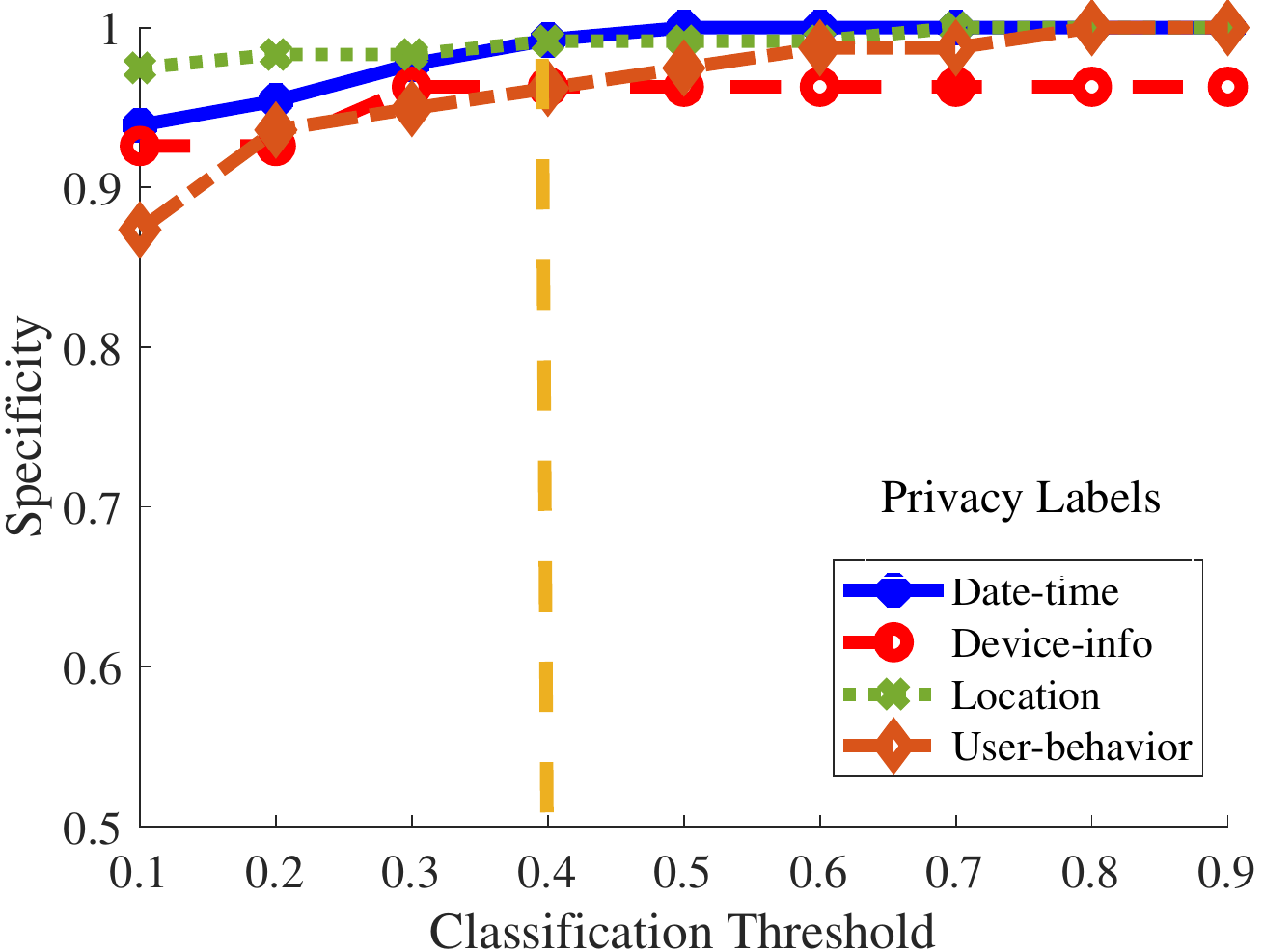}\label{fig:specificty}}
    \subfigure[Average Value]{\includegraphics[width=0.205\textwidth]{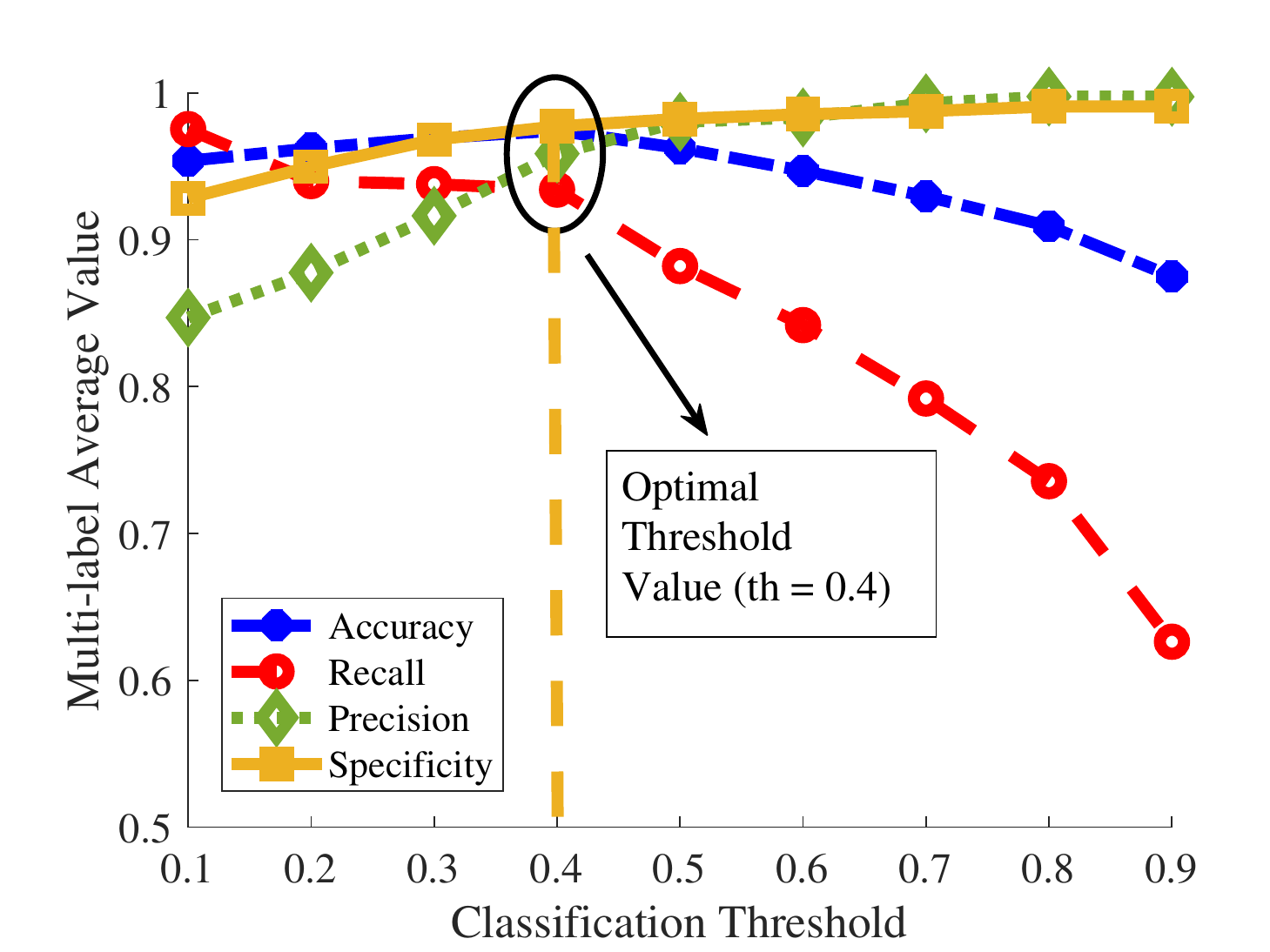}\label{fig:ave_metrics}}
    \vspace{-0.1in}
    \caption{Evaluation results in the classification of IoT strings extracted from messaging and Internet communications to user-friendly privacy labels: (a) accuracy, (b) recall, (c) precision, (d) specificity, and (e) the average value of all considered performance metrics.}
    \label{fig:metrics}
    \vspace{-0.2in}
\end{figure*}

\section{Evaluation}
\label{sec:evaluation}
We evaluated \daint based on the three research questions (RQ) below:
\begin{itemize}[wide=0pt]
    \setlength\itemsep{0.0em}
    \item[\textbf{RQ1}] What is the effectiveness of \daint in correctly classifying the IoT strings into privacy labels? (Section~\ref{sec:evaClassification}).
    \item[\textbf{RQ2}] What is the effectiveness of \daint in detecting sensitive data leaks and potential privacy behaviors in IoT apps? (Section~\ref{sec:evaAnalyzer}).
    \item[\textbf{RQ3}] What is the runtime performance overhead of \daint in terms of latency and storage? (Section~\ref{sec:overhead}).
\end{itemize}

We used a total of 540 current IoT apps to implement and evaluate \daint. Out of the total, 380 \smarthings apps were used to build the IoT corpus and train the NLP model, and the other 160 apps were used to evaluate its performance. For comprehensiveness, we included in the evaluation phase 120 \smarthings market apps~\cite{Official, Community}, and 40 malicious apps crawled from the \iotbench repository~\cite{iotbench-repo}. \iotbench is an IoT-specific test corpus used to evaluate systems designed for IoT app security and privacy. It includes flawed apps that perform various malicious activities, including sensitive data leaks via both messaging and Internet functions. After instrumentation of the evaluation apps was completed, we executed the instrumented apps in the \smarthings IDE~\cite{simulator}, which is a propriety simulation environment to execute \smarthings apps. \daint's instrumentor adds 25\% more lines of code (LoC) to the apps on average, which translates into adding 65 LoC to an IoT app that has an average size of 265 LoC. While being executed in \smarthings IDE, the instrumented apps send their privacy data to \daint's analyzer which runs on a Python web server hosted on Google App Engine~\cite{googleengine}. 

We evaluated \daint's accuracy in classifying strings extracted from IoT messaging and Internet communications into users' privacy preferences (i.e., privacy labels) (Section~\ref{sec:evaClassification}). Also, we check its effectiveness in detecting sensitive data leaks and flagging potential privacy behaviors in IoT apps (Section~\ref{sec:evaAnalyzer}). As noted earlier in Section~\ref{sec:terms}, a leak is defined as a piece of information that leaves an IoT app without user consent. In IoT apps, the consent of a user is either asked through an app description block or through acknowledging the flow's recipients at install-time. However, we found that, out of 160 apps, only 121 (75.6\%) define some type of app's description. Additionally, in most cases users are not informed nor given the chance to assert the app's intent or the recipients of sensitive information. For instance, a messaging function using a hard-coded recipient not entered by a user, or an Internet communications that sends sensitive data to a server that is not informed to the user via an app description block is clearly a leak. Finally, we define privacy behaviors that put the sensitive information at risk. For instance, an Internet communication that sends data to a remote server without using encryption.

\subsection{Performance of IoT String Classification}
\label{sec:evaClassification}
\daint's analyzer classifies IoT app strings into four privacy labels. The classification results include the assigned label and the confidence scores through a threshold $th$. If the classification score is over the predefined threshold, the privacy label is assigned to the string. For instance, the string ``the front door at home is unlocked'' resulted in classification scores of device-info=0.94, user-behavior=0.03, location=0.86, and date-time=0.3. By design, the classification scores are independent of each other, that is, if $th$ value is set to 0.7, \daint classifies the string into privacy information of type device-info and location. In total, \daint classified 146 IoT strings extracted from 95 different IoT apps. Out of these, 112 strings were extracted from messaging communications; 54 messages from 44 market IoT apps and 58 messages from 30 malicious apps. The remaining 34 strings were extracted from Internet communications; 12 Internet calls extracted from 10 market IoT apps, and 22 extracted from 11 malicious IoT apps.

\vspace{1pt}\noindent\textbf{Classification of Encrypted IoT Strings.} As detailed in Section~\ref{sec:selective}, the selective instrumentation of \daint enables NLP analysis on encrypted data. We study the effectiveness of \daint in classifying IoT strings that have been previously encrypted. Out of 160 apps analyzed, we found seven benign apps executing nine encrypted Internet communications. Similarly, we found 11 malicious apps executing 23 Internet communications with encrypted content. Despite the use of encryption techniques, \daint effectively collected and classified the information sent as encrypted strings in 100\% of the cases. Finally, we did not find any IoT application using encrypted messaging.

\vspace{1pt}\noindent\textbf{Accuracy by Threshold Values.} We study how \daint's classifier performs for different threshold values $th$. The goal of this analysis is to find the value $th$ that leads to the highest performance overall. Figure~\ref{fig:metrics} illustrates the accuracy, recall, precision, and specificity of the NLP model. Overall, \daint yields the best accuracy with the threshold values between 0.1 to 0.5. Figure~\ref{fig:ave_metrics} summarizes the average metrics for different values of $th$. We observe that $th$ = 0.4 yields the best classification results for all metrics. Finally, \daint classifies IoT strings to correct privacy labels with 94.25\%, 85.17\%, 95.01\%, and 97.34\% average accuracy, recall, precision, and specificity, respectively. We detail the impact of various threshold values to classification results in Appendix~\ref{sec:threshold}. 

\begin{table*}[!t]
\centering
{\footnotesize{%\addtolength{\tabcolsep}{-3pt}
\resizebox{\textwidth}{!}{
\setlength{\tabcolsep}{15pt}
\begin{threeparttable}
\begin{tabular}{|c|c|c|c|c|c|c|}
\hline
%\textbf{\begin{tabular}[c]{@{}c@{}}No.\end{tabular}} &
\textbf{\begin{tabular}[c]{c}Type\textsuperscript{*}\end{tabular}} &
\textbf{\begin{tabular}[c]{@{}c@{}}App Name\end{tabular}} &
\textbf{\begin{tabular}[c]{@{}c@{}}Leak Type\textsuperscript{**}\end{tabular}} & \textbf{\begin{tabular}[c]{@{}c@{}}Recipient\end{tabular}} &
\textbf{\begin{tabular}[c]{@{}c@{}}Content\end{tabular}} & \textbf{\begin{tabular}[c]{@{}c@{}}Behavior\end{tabular}} &
\textbf{\begin{tabular}[c]{@{}c@{}}Privacy Labels\\\end{tabular}} \\ \hline \hline
\rowcolor{light-gray}
\textbf{\begin{tabular}[c]{c}\multirow{8}{*}{\circled{1}}\end{tabular}}  & Squeeze Box Controller~\cite{Community}  &  \texttt{I} & http://\$ip:\$port  & ``Open box''  & \bcircle  & Device-info  \\
&  StatHat Quick Start~\cite{Community}  &   \texttt{I}  & http://api.stathat.com/ez  & Thermostat1:thermostat & \bcircle & Device-info \\
\rowcolor{light-gray}
& ThingSpeak Logger~\cite{Community}   &  \texttt{I}  &  http://api.thingspeak.com & DeviceID:$333$:Key:$123$  & \bcircle  & Device-info  \\
&  \begin{tabular}[c]{@{}c@{}} User Lock Manager~\cite{Community}  \end{tabular}&   \texttt{M}  & defined by user  & \begin{tabular}[c]{@{}c@{}} User no longer has \\access to the door \end{tabular}& \wcircle  & \begin{tabular}[c]{@{}c@{}}Device-info\\User-behavior  \end{tabular}\\
\rowcolor{light-gray}
& \begin{tabular}[c]{@{}c@{}} Smart Lock~\cite{Community}  \end{tabular} &   \texttt{M}  & defined by user  & \begin{tabular}[c]{@{}c@{}} SmartLock disabled. \\Door will remain \\unlocked indefinitely \end{tabular}& \wcircle  & \begin{tabular}[c]{@{}c@{}}Device-info\\User-behavior  \end{tabular} \\ \hline \hline
\textbf{\begin{tabular}[c]{c}\multirow{9}{*}{\circled{2}}\end{tabular}} &  Fire Alarm~\cite{iotbench-repo}   &   \texttt{I}  & \begin{tabular}[c]{@{}c@{}}http://$141.212.110.244$/stmalware/ \\ maliciousServer.php\end{tabular} & -- & \bcircle  & --  \\
\rowcolor{light-gray}
&  \begin{tabular}[c]{@{}c@{}}  Ransomware~\cite{iotbench-repo} \end{tabular}  & \texttt{I}  & \begin{tabular}[c]{@{}c@{}}http://$141.212.110.244$/stmalware/ \\ maliciousServer.php\end{tabular}  & --  & \bcircle  & --  \\
&  \begin{tabular}[c]{@{}c@{}}Remote Command~\cite{iotbench-repo}\end{tabular}  & I  & \begin{tabular}[c]{@{}c@{}}http://$141.212.110.244$/stmalware/ \\ maliciousServer.php\end{tabular}  & --  & \bcircle  & -- \\
\rowcolor{light-gray}
&  Spyware~\cite{iotbench-repo}  &  \texttt{M}  & $123-456-7890$  & \begin{tabular}[c]{@{}c@{}}Doors locked after\\ everyone departed\end{tabular}  & \bcircle  & \begin{tabular}[c]{@{}c@{}}Device-info\\User-behavior  \end{tabular}\\
 &  \begin{tabular}[c]{@{}c@{}} User Event~\cite{iotbench-repo}  \end{tabular} &  \texttt{M} & $123-456-7890$  & \begin{tabular}[c]{@{}c@{}} Everyone is away\\ and Hub ID is $123$ \end{tabular}& \bcircle  & \begin{tabular}[c]{@{}c@{}}Device-info\\Location\\User-behavior  \end{tabular}\\\hline
\end{tabular}
\begin{tablenotes}
    \item * \circled{1} is for Market IoT apps and \circled{2} is for Malicious IoT apps (Handcrafted)\\
    ** \texttt{I} is for Internet and \texttt{M} is for Messaging.
    \bcircle~~is for Privacy Risk and \wcircle~~is for Privacy Preference of a user.
\end{tablenotes}
\end{threeparttable}
}
}}
\vspace{-0.1in}
\caption{Examples of privacy risks and the use of sensitive information in market and malicious IoT apps. \daint identified six privacy-violating behaviors and 62 leaks from market and malicious IoT apps.}
\vspace{-0.1in}
\label{tab:concerns}
\end{table*}

\begin{table}[!t]
\centering
%\footnotesize
\setlength{\tabcolsep}{4pt}
\resizebox{\columnwidth}{!}{
\begin{threeparttable}
\begin{tabular}{|ccccc>{\columncolor[gray]{0.9}}c|}
\hline
\textbf{\begin{tabular}[c]{@{}c@{}}App \\ Type\end{tabular}} & 
\textbf{\begin{tabular}[c]{@{}c@{}}Total\\ Apps \\Analyzed\end{tabular}} &
\textbf{\begin{tabular}[c]{@{}c@{}}Messaging \\Ccommunications \\Analyzed\end{tabular}} & \textbf{\begin{tabular}[c]{@{}c@{}}No. of \\Data\\ Leaks \end{tabular}}  & 
\textbf{\begin{tabular}[c]{@{}c@{}}No. of \\Privacy\\ Concerns  \end{tabular}} &
\textbf{\begin{tabular}[c]{@{}c@{}}\daints \\ Effectiveness\end{tabular}} \\ \hline
\hline
%\rowcolor{light-gray}
\textbf{Market}     & 120 & 54 & 0 & --  & -- \\ 
\textbf{Malicious}  & 40  & 58 & 29  & --  & 100\%  \\ \hline
\rowcolor{light-gray}
\textbf{Total}  & 160  & 112 & 29  & --  & 100\%  \\ \hline
\end{tabular}
\end{threeparttable}
}
\vspace{-0.10in}
\caption{Effectiveness of \daint in detecting sensitive data leaks using messaging.}
\vspace{-0.1in}
\label{tab:context_MSG}
\end{table}

\begin{table}[!t]
\centering
%\scriptsize
\setlength{\tabcolsep}{4pt}
\resizebox{\columnwidth}{!}{
\begin{threeparttable}
\begin{tabular}{|ccccc>{\columncolor[gray]{0.9}}c|}
\hline
\textbf{\begin{tabular}[c]{@{}c@{}}App \\ Type\end{tabular}} & 
\textbf{\begin{tabular}[c]{@{}c@{}}Total\\ Apps \\Analyzed\end{tabular}} &
\textbf{\begin{tabular}[c]{@{}c@{}} Internet\\ Communications\\Analyzed\end{tabular}} & 
\textbf{\begin{tabular}[c]{@{}c@{}}No. of \\Data\\ Leaks\end{tabular}}  & 
\textbf{\begin{tabular}[c]{@{}c@{}}No. of \\Privacy\\ Concerns \end{tabular}} & 
\textbf{\begin{tabular}[c]{@{}c@{}}\daints \\ Effectiveness\end{tabular}} \\  \hline
\hline
%\rowcolor{light-gray}
\textbf{Market}     & 120 & 12 & 11 & 3\textdagger & 100\% \\ 
\textbf{Malicious}  & 40  & 22 & 22 & 3\textdaggerdbl & 100\% \\ \hline
\rowcolor{light-gray}
\textbf{Total}  & 160  & 34 & 33 & 6 & 100\% \\ 
\hline
\end{tabular}
\begin{tablenotes}
    \item[\textdagger] Includes one privacy concern that does not constitute a data leak and two privacy concerns that were also flagged as data leaks.
    \item[\textdaggerdbl] Includes three privacy concerns that were also flagged as data leaks.
\end{tablenotes}
\end{threeparttable}
}
\vspace{-0.10in}
\caption{Effectiveness of \daint in detecting sensitive data leaks via Internet communications.}
\vspace{-0.20in}
\label{tab:context_INT}
\end{table}

\vspace{1pt}\noindent\textbf{Accuracy by Privacy Label.} We further study the sensitivity of \daint's classifier to each privacy label. The goal is to determine the effectiveness of \daint in classifying strings to the different privacy labels. \daint achieves the highest accuracy for privacy labels of type location and date-time. This is because date, time, and location can be easily inferred from semantically-simple strings. Also, it is very common to find information related to these privacy labels embedded in the same string (e.g., ``he arrived home 5 minutes ago''). In contrast, \daint obtained the lower accuracy results for privacy labels of type user-behavior. This is because user-behavior information is harder to infer from simple strings. In spite of these results, \daint achieved the lowest accuracy of 90.79\% for user-behavior labels, which is comparable with the best classification results of other similar tools in the market~\cite{flowcog} (more details in Appendix~\ref{sec:threshold}). 

\vspace{1pt}\noindent\textbf{Findings on Privacy Analysis of IoT Strings.} \daint classifies IoT strings to privacy labels with an average accuracy of 94.25\%. Out of 160 apps, \daint identified 50 (31.25\%) apps that transmit data related to device information via messaging, and 20 (12.5\%) apps that do the same via Internet communications. Also, 11 (6.9\%) apps handled data related to date and time in messages and only one transmit similar information using the Internet. \daint also identified 38 (23.75\%) apps transmitting information related to the user behavior in their messages and nine (5.6\%) including similar type of information in Internet communications. Further, 20 (12.5\%) apps sent information related to location via messaging, while six (3.7\%) apps did the same via Internet. Finally, we evaluated how the privacy analysis of IoT strings benefited from the use of NLP. \daint assigned multiple privacy labels to classify more semantically-complex IoT strings, which guaranteed completeness in the privacy analysis. Out of 146 strings analyzed, \daint applied multi-labeling to 68 (46.5\%). Specifically, 54 IoT strings (36.9\%) received two privacy labels and 14 (9.6\%) strings received three privacy labels. Additional details on the classification of IoT strings to privacy labels are provided in Appendix~\ref{sec:labels}. 

\subsection{Analysis of Data Leaks in IoT Apps}
\label{sec:evaAnalyzer}
We evaluated the correctness of \daint in identifying data leaks. Table~\ref{tab:concerns} presents examples of data leaks in IoT apps, and the associated privacy labels assigned to the leak. 

\vspace{1pt}\noindent\textbf{Findings on Data Leaks via Messaging.} Table~\ref{tab:context_MSG} shows \daint's findings after analyzing messaging recipients. \daint extracted 54 recipients in messages from market apps and 58 from malicious apps. We found no data leaks via messaging in market apps, meaning, all recipients were defined (or authorized) by the user at install-time. We believe this is due the strict review process enforced by \smarthings IoT market. Further, we found 29 leaks from 14 different malicious apps~\cite{iotbench-repo}, meaning, all these recipients were hard-coded by a developer and their intent were not defined in the app's description block. For instance, the User Event app~\cite{iotbench-repo} leaks privacy labels of type device-info, location, and user-behavior (``Everyone is away and hub ID is \#'') to a hard-coded phone number. We manually reviewed the app source codes and verified that all \daint's findings were correct. Also, we verified that all data leaks via messaging were properly flagged.  

\vspace{1pt}\noindent\textbf{Findings on Data Leaks via Internet.} Table \ref{tab:context_INT} details the effectiveness of \daint's analyzer in finding data leaks via Internet communications. \daint analyzed 34 recipients from Internet communications, 12 from market and 22 from malicious apps. Our tool flagged 11 Internet communications from seven market apps that leak privacy data without user consent. That is, the user is neither informed about the recipient of the data in the app description block nor enters the URL or domain name herself. For instance, the ThingSpeak Logger app~\cite{Community} transmits a device ID to a remote server through an HTTP call, which is identified through a device-info privacy label. For malicious apps, \daint flagged 22 Internet communications from 14 different malicious apps as leaks. We manually verified that all \daint's findings were correct. Also, we verified that 100\% of data leaks via Internet were properly flagged.  
\begin{table*}[!t]
\centering
\resizebox{\textwidth}{!}{
\setlength{\tabcolsep}{0.165in}
\renewcommand{\arraystretch}{1}
{\scriptsize{
\begin{tabular}{|lccccccccc|}
\hline
\textbf{\begin{tabular}[c]{@{}c@{}}Tool \\Name\end{tabular}} &
\textbf{\begin{tabular}[c]{@{}c@{}}Domain\end{tabular}} & \textbf{\begin{tabular}[c]{@{}c@{}}Source Code\\ Analysis\end{tabular}} & \textbf{\begin{tabular}[c]{@{}c@{}}Dynamic\\ Analysis\end{tabular}} &
%\textbf{\begin{tabular}[c]{@{}c@{}}Context\\ Analysis\end{tabular}} &
\textbf{\begin{tabular}[c]{@{}c@{}}Semantics\\ Analysis\end{tabular}} &
\textbf{\begin{tabular}[c]{@{}c@{}}Privacy\\ Analysis\end{tabular}} &
\textbf{\begin{tabular}[c]{@{}c@{}}NLP/ML\\ Analysis\end{tabular}} &
\textbf{\begin{tabular}[c]{@{}c@{}}User\\ Awareness\end{tabular}} &
\textbf{\begin{tabular}[c]{@{}c@{}}Overhead\\ Evaluation\end{tabular}} &
\textbf{\begin{tabular}[c]{@{}c@{}}Freely\\ Available\end{tabular}} \\ \hline
%\textbf{\multicolumn{1}{c}{\begin{tabular}[c]{@{}l@{}}Comments\end{tabular}}} \\ \hline
\hline \hline
\rowcolor{light-gray}
TaintDroid~\cite{EnckTaintDroid} & Android  & \bcircle  & \bcircle  & \wcircle  &  \bcircle & \wcircle  & \bcircle  & \bcircle & \bcircle   \\
FlowDroid~\cite{ArztFlowdroid} & Android  & \bcircle  & \wcircle   & \wcircle & \bcircle  & \wcircle & \wcircle  & \wcircle & \bcircle   \\
\rowcolor{light-gray}
FlowCog~\cite{flowcog} & Android  & \bcircle  & \bcircle  & \bcircle &  \wcircle & \bcircle & \bcircle  & \wcircle  &   \bcircle  \\ 
FlowFence~\cite{fernandes2016flowfence}  & IoT  & \bcircle  & \bcircle  & \wcircle &  \wcircle   & \wcircle & \wcircle   &  \bcircle  & \bcircle  \\
\rowcolor{light-gray}
ContextIoT~\cite{jia2017contexiot} & IoT  & \bcircle  & \bcircle  & \wcircle &  \wcircle & \wcircle & \bcircle  &  \bcircle  &  \wcircle   \\
\saints~\cite{saint}   & IoT  & \bcircle  & \wcircle  & \wcircle &  \wcircle  & \wcircle & \bcircle  &  \wcircle  & \bcircle \\
\rowcolor{light-gray}
ProvThings~\cite{fear} & IoT  & \bcircle  & \bcircle  & \wcircle &  \wcircle & \wcircle&  \wcircle  &  \bcircle  & \wcircle  \\
iRuler~\cite{iruler} & IoT  & \bcircle  & \wcircle  & \bcircle &  \wcircle & \bcircle&  \wcircle  &  \wcircle  & \wcircle  \\
\rowcolor{light-gray}
\daints   & IoT  & \bcircle & \bcircle  & \bcircle &  \bcircle & \bcircle & \bcircle  &  \bcircle  & \bcircle   \\
\hline
\end{tabular}
}}}
\vspace{-0.1in}
\caption{Comparison between \daint and other similar analysis tools for Android and IoT apps.}
\vspace{-0.1in}
\label{tab:IFA}
\end{table*}

\vspace{1pt}\noindent\textbf{Findings on Privacy Behaviors via Internet.}
\daint further verifies whether an IoT app protects the sensitive information before sending it to an external party. The use of encryption gives the sensitive data extra protection against unauthorized disclosure to passive eavesdroppers. Out of 12 Internet communications analyzed from market apps, \daint flagged three from three different apps which make unsecured \texttt{HTTP} requests. Interestingly, one of these calls was authorized by the user at install-time through the app description block. Coincidentally, our tool also flagged three Internet communications from three different malicious apps that constituted potential privacy behaviors. Finally, \daint did not detect any privacy behavior via messaging. 

\subsection{Overhead Analysis}
\label{sec:overhead}
We evaluated the performance overhead of \daint in terms of runtime and storage overhead.

\vspace{1pt}\noindent \textbf{Runtime Overhead.} Latency refers to the time elapsed from the moment the IoT app sends relevant data via \daint's API (see Section~\ref{sec:api}) to the moment that the user receives \daint's notifications. Latency overhead is calculated as the average difference in the execution time of the original and instrumented apps. On the one hand, \daint required 75 ms to perform the NLP-based classification of the sink-call contents on average. On the other hand, we implemented \daint's API using asynchronous \texttt{HTTPS} communications via the \texttt{$asynchttp\_v1$} class defined in \smarthings~\cite{smartThings-documentation}. With this, we obtained a communication latency of 35 ms on average while supporting encrypted communications that protected the sensitive information sent to \daint's server. In the end, we found that the total latency introduced by \daint was 105 ms on average. 

\vspace{1pt}\noindent\textbf{Storage Overhead.} We measured the storage overhead imposed by \daint. Our tool does not store app information after the analysis is completed; thus, the storage cost is determined by the total storage size of the JSON object used to exchange information between the IoT apps and \daint's analyzer (Section \ref{sec:implementation}). We evaluated the storage overhead imposed by the analysis of 160 IoT apps. On average, \daint imposes 1 KB of storage overhead, which we consider negligible.  
\vspace{-0.15in}

%%%%%%%%%%%%%%%%%%%%%%%%%%%%%%%%%%%%%%%%%%%%%%%%%%%%%%%%%%%%%%%%%%%%%%%%%%%%%%%%%%%%%%%%%%%%%%

\section{Discussion}
\label{sec:benefits}

\daint is the first dynamic tool that performs NLP-based real-time privacy analysis in IoT apps to (1) classify IoT strings to privacy labels that are easy to understand by the user, and to (2) flag IoT apps that represent privacy concerns for the user. We implemented \daint for SmartThings IoT platform, and we plan to extend our analysis to other IoT platforms. Additionally, we analyzed 380 IoT apps and constructed a dataset to study how these apps use privacy-sensitive information. While our corpus included IoT strings extracted from SmartThings market apps, we plan to investigate other IoT platforms to construct similar datasets. 

We designed and built \daint by first understanding the privacy needs of IoT users. We plan to conduct an additional study to evaluate the usability of \daint which is outside the scope of the current work. Also, \daint's analysis would benefit from mapping the app descriptions to privacy labels. However, this is challenging as the description block of an IoT app does not explicitly state the app's privacy behavior but its functionality. We plan to use more advanced NLP techniques to address this challenge. Finally, \daint's execution requires the collection and analysis of privacy-sensitive information. We use secure \texttt{HTTPS} communications to protect the communication between IoT apps and \daint's server. In addition, \daint does not keep record of any collected information nor share this information with any third party. As a future work, a complete privacy assessment of \daint may be conducted to guarantee that user's privacy is completely preserved. 

%%%%%%%%%%%%%%%%%%%%%%%%%%%%%%%%%%%%%%%%%%%%%%%%%%%%%%%%%%%%%%%%%%%%%%%%%%%%%%%%%%%%%%%%%%%%%%%%%%%%

\section{Related Work}
\label{sec:relatedwork}
Information Flow Analysis (IFA) tools have been proposed for the mobile phone \cite{EnckTaintDroid, ZhuTaint, GuTaint, clause2007dytan, ArztFlowdroid, gordon2015information} and IoT apps \cite{fernandes2016security, saint, rahmati2016applying, iotdots}. Table \ref{tab:IFA} compares \daint with current IFA tools. Though these solutions are effective in identifying sensitive data flows, most of them either consider security risks from data flows with tainted variables or via inter-rule vulnerabilities. FlowCog~\cite{flowcog} recently proposed a solution to address these limitations. Here, the authors establish data flow dependencies based on Android app view context, which cannot be extracted from IoT apps due to specific architectural differences.  

There exist a few systems for IoT app data flow analysis. FlowFence~\cite{fernandes2016flowfence} makes the consumers of IoT data declare their intended activities to enforce data flow policies. More recently, \saint, a static analysis tool, is proposed to detect sensitive data flows in IoT apps~\cite{saint}. \saint does not consider sensitive data leaks at runtime and data leaks-through strings, while FlowFence often over-approximates the data leaks. Indeed, our analysis showed that 64\% of the analyzed apps leak sensitive data through simple strings that do not include information from taint variables, yet the strings itself is sensitive; thus the analysis of strings is required. iRuler~\cite{iruler} applies NLP techniques to uncover inter-rule vulnerabilities parsed not from apps, but IoT services. Similarly, the authors in~\cite{ifttt} analyze privacy risks in IoT online recipes. Lastly, ProvThings~\cite{fear} uses static and dynamic analysis to collect data provenance and identify the root cause of attacks in IoT apps. However, this work is limited to analyze dependencies between events and data states and does not offer any built-in privacy analysis. %\daint effectively combines the app source code instrumentation, dynamic analysis of data flows, and NLP techniques to uncover privacy risks in IoT apps, and allows users to specify their privacy preferences at install time. 

\vspace{1pt}\noindent \textbf{Differences from Existing Works.} \daint is a novel dynamic tool that uncovers privacy risks of IoT apps. Considering the privacy expectations of IoT users, \daint instruments IoT apps to collect and analyze the data sent to external parties in real time. \daint performs rich NLP-based classification of IoT strings to four user-friendly privacy labels, which are also customizable. Also, \daint analyzes the recipients of the sensitive information to uncover data leaks and privacy behaviors. Finally, \daint implements notification mechanisms to inform its findings to the user.   
\vspace{-0.15in}

%%%%%%%%%%%%%%%%%%%%%%%%%%%%%%%%%%%%%%%%%%%%%%%%%%%%%%%%%%%%%%%%%%%%%%%%%%%%%%%%%%%%%%%%%%%%%%%%%%%%%%

\section{Conclusions}
\label{sec:conclusion}
IoT apps access sensitive data that, if leaked, can compromise the privacy of the users. IoT platforms do not offer real-time privacy analysis that informs users about how the IoT apps use sensitive information and what they do with it. In this paper, we introduced \daint, a novel dynamic privacy analysis tool for current IoT apps. We designed and built \daint based on a privacy survey with 123 IoT users. \daint uses NLP techniques to classify IoT strings extracted from messaging and Internet communications to a set of privacy labels at runtime. Additionally, \daint enables users to select their privacy preferences and reports its findings based on their privacy profiles. This allows users to make informed decisions about their privacy and reject apps. We analyzed 540 real IoT apps to train the NLP model and evaluate its effectiveness. \daint classifies IoT strings to correct privacy preferences with an average accuracy of 94.25\% and flags 35 apps that leak data. Finally, \daint imposes minimal overhead to an IoT app's execution, introducing on average 105 ms additional latency.

%%%%%%%%%%%%%%%%%%%%%%%%%%%%%%%%%%%%%%%%%%%%%%%%%%%%%%%%%%%%%%%%%%%%%%%%%%%%%%%%%%%%%%%%%%%%%%%%%%%%%

\section{Acknowledgments}
\label{sec:ack}
This material is based upon work partially supported by the U.S. National Science Foundation under award numbers NSF-CAREER-CNS-1453647 and NSF-1663051.

%%%%%%%%%%%%%%%%%%%%%%%%%%%%%%%%%%%%%%%%%%%%%%%%%%%%%%%%%%%%%%%%%%%%%%%%%%%%%%%%%%%%%%%%%%%%%%%%%%%%

%\input{Introduction}

%\input{UserStudy}

%\input{Problem}

%\input{Architecture}

%\input{Implementation}

%\input{Evaluation}

%\input{Discussion}

%\input{RelatedWork}

%\input{Conclusions}

%\input{ACK}

{\footnotesize\bibliographystyle{acm}

}

\setcounter{figure}{0}
\setcounter{table}{0}
\setcounter{lstlisting}{0}
\setcounter{section}{0}
\setcounter{equation}{0}

\appendix
%\newpage \clearpage

\vspace{0.02in}
\appendix{}
\label{sec:appendix}

\setcounter{figure}{0}
\setcounter{table}{0}
\setcounter{lstlisting}{0}
\setcounter{section}{0}

\section{Example IoT Privacy Survey Questions}
\label{sec:examplesurvey}
We present a list of representative IoT privacy survey questions from all the categories. The entire user study can be found at \texttt{https://anonymous.com}.
\vspace{-0.1in}
\subsection{User Characterization}

\begin{enumerate}
    \item {Do you use, have used, or are you planning to use any IoT device?\\
    ( ) Yes\\
    ( ) No\\
    ( ) Maybe}\\
   
    \item{What is your technical experience with IoT apps?\\
    ( ) I can build, implement, code my own IoT app\\
    ( ) Installed/can install/configure an IoT app using the source code available online\\
    ( ) Installed/can install/configure an IoT app's marketplace (Google Play, App Store, etc.)\\
    ( ) I just know how to press the buttons\\
    ( ) I have no idea how to deal with IoT apps}
\end{enumerate}

\vspace{-0.1in}
\subsection{Security and Privacy Concerns in Smart Apps}

\begin{enumerate}
    \item {What information would you consider sensitive if  used in IoT apps/devices? Please check all that apply.\\
    ( ) My personal data (e.g., email address, phone number, residential address, etc.)
    ( ) Whatever I do, my behavior (e.g., when I arrive home, when I leave home, I go to sleep, etc.)\\
    ( ) My (or my devices') location (e.g., my location while using the apps, etc.)\\
    ( ) My device settings/the way I configure the devices (e.g., Time of the day the lights turn On/Off, my thermostat temperature settings, etc.)\\
    ( ) My (or my devices') timing (e.g., the time passed since I left home, the time passed since I went to sleep, etc.)\\
    ( ) Information from my devices (e.g., device type, manufacturer, device IDs, etc.)\\
    ( ) Data from my devices (e.g., door state open or close, light on or off, etc.)\\
    ( ) Other \\ \\
    If you selected "Other", please explain: \\
    }

    \item {Have you heard or personally have privacy concerns on the use of the IoT devices and systems?\\
    ( ) Big concerns
    ( ) Some concerns\\
    ( ) I do not, but I know someone that does\\
    ( ) Never thought about it, until now\\
    ( ) I do not care
    }
\end{enumerate}

\begin{figure*}[!t]
    \centering{\includegraphics[width=1\textwidth]{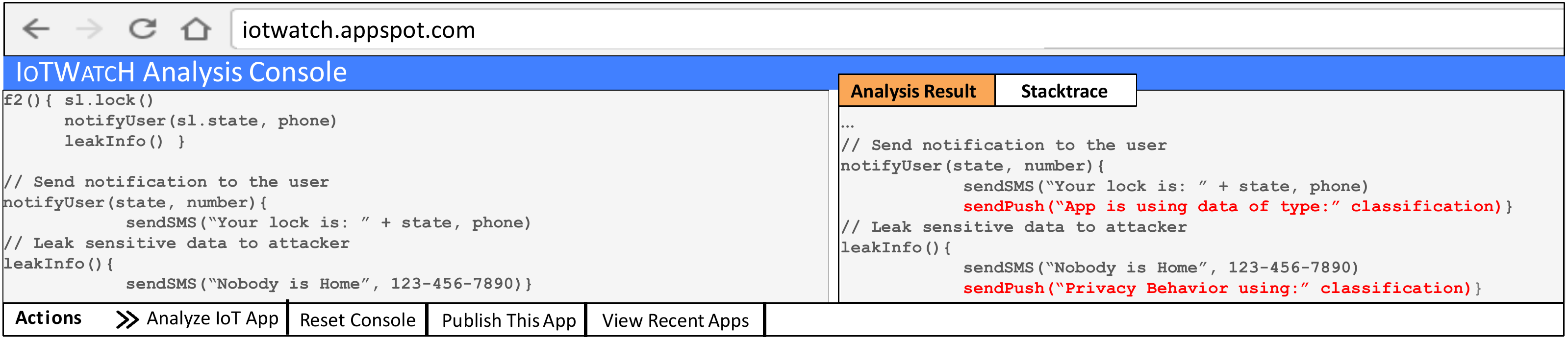}}
    \caption{The left console is the analysis area where the user inputs the original IoT app. The right console returns the output of the instrumentation process. We made \daint's instrumentor freely available to the community at \texttt{https://\daint.appspot.com/}.}
    \vspace{-0.2in}
    \label{fig:webtool}
\end{figure*}

\subsection{Privacy Analysis Tools and Features}

\begin{enumerate}
    \item {Do you think there is a need for a tool to check for security privacy risks from the smart apps?\\
    ( ) Yes\\
    ( ) No\\
    ( ) Maybe\\ }
    \item{Would you be willing to use available automatic tools that analyze and modify smart apps to enable security and privacy analysis in real-time?\\
    ( ) Yes\\
    ( ) No\\
    ( ) Maybe\\ }
    
    %\item {Assume that you wish classify the type of data collected/used by smart apps into different categories so that they are easily understood by any level of users/consumers. Would you use the category  "Device Info" to identify information related to the devices (e.g., device type, branch, device ID, etc.)? \\
    %( ) Yes\\
    %( ) No\\
    %( ) I have a better category suggestion\\ \\
    %If you answered "I have a better category suggestion", please share your idea below:\\
    %}
\end{enumerate}
\vspace{-0.2in}
\section{More Details on \daint's Instrumentor} 
\label{sec:algorithm}
In Appendix~\ref{sec:algorithm}, we detail Algorithm~\ref{algo:instrumenter}, which describes the source code analysis and app instrumentation process in \daint. Our privacy tool requires open access to the IoT apps' source code (Line \ref{line:ini1}). This requirement does not constitute a limitation since several IoT platforms (e.g., \smarthings, OpenHAB) make the source code of the apps available online through official and third-party repositories \cite{Official, openHABMarket, Community}. The first step towards analyzing the apps is to generate the IR (Line \ref{line:ast}) and extract the ICFG (Line \ref{line:icfg}) of the app. From there, the app's data-flow analysis starts by flagging relevant app and data-flow information (e.g., user-defined data, data-flow recipients). Then, the automatic instrumentation process starts by inserting extra code that defines the different \daint-related $global$ variables inside the \texttt{installed()} and \texttt{updated()} methods (Lines \ref{line:state1}, \ref{line:state2}, \ref{line:state3}, and \ref{line:state4}). In that way, \daint always keeps the current value of the flagged data through all the different node executions. Also, the instrumentor implements a UI to define the privacy preferences of the user (Line~\ref{line:UI}). Finally, one can notice from Line \ref{line:sms} and Line \ref{line:int} that two different methods are inserted to monitor messaging and Internet communications, respectively. 

\begin{algorithm}[ht!]
\footnotesize
 \caption{\textit{App source code analysis and code instrumentation in \daint}}
 \label{algo:instrumenter}
 \begin{algorithmic}[1]

\STATE $appSC \gets$ app source code\\                            \label{line:ini1}
\COMMENT {Source Code Analysis}
\STATE $IR \gets$ generateIR($appSC$)\\                         \label{line:ast}
\STATE $ICFG \gets$ generateICFG($appSC$)\\                       \label{line:icfg}               
\COMMENT{Instrumentation}
\FOR{All $ICFG$ nodes invoking $sinks$} 
    \IF{$node$ has not been previously visited}
        \IF{$node$ invokes $install$ method}                             
            \STATE $global \gets$ user-defined recipients            \label{line:state1}
            \STATE $global \gets$ user-defined URLs                  \label{line:state2}
            \STATE Implements Privacy Settings UI                    \label{line:UI}
            \STATE Flag $ICFG$ node as visited 
        \ENDIF
        \IF{$node$ invokes $install$ method}                             
            \STATE $global \gets$ user-defined recipients            \label{line:state3}
            \STATE $global \gets$ user-defined URLs                  \label{line:state4}
            \STATE Flag $ICFG$ node as visited 
        \ENDIF
        \IF {$node$ invokes $update$ method}                             
            \STATE Update $global$                                   \label{line:update} 
            \STATE Flag $ICFG$ node as visited 
        \ENDIF
        \IF {$node$ invokes messaging $sink$}                             
            \STATE Insert $WatchMsg$ method                         \label{line:sms}
            \STATE Flag $ICFG$ node as visited 
        \ENDIF
        \IF {$node$ invokes Internet $sink$}                                 
            \STATE Insert $WatchInt$ method                         \label{line:int}
            \STATE Flag $ICFG$ node as visited 
        \ENDIF
    \ENDIF
\ENDFOR
\end{algorithmic}
\end{algorithm}

\section{\daint's Instrumentor Online}
\label{sec:web}
We made the instrumentor available online at: \texttt{https://\daint.appspot.com/}. Figure \ref{fig:webtool} depicts details of the online version of \daint's instrumentor. At the left console, the user inputs the IoT app source code that needs to be modified to enable \daint, and at the right console, the tool automatically returns the \daint-instrumented app. Below, we detail the implementation steps of \daint.

\section{Additional Details on \daint Analyzer}
\label{sec:additionalanalyzer}
Algorithm \ref{algo:analytics} details the steps followed in \daint's analyzer. First, in Line \ref{line:api}, the analyzer extracts all the information received from the IoT app through \daint's REST API. Also, variables are initialized in Lines from \ref{line:inistart} to \ref{line:iniend}. After \daint completes the initialization step, the tool applies multi-class classification on the app's data-flow contents in Line \ref{line:nlp}. Then, it also performs data-flow recipient analysis by correlating recipients from sink-calls with user-defined data to detect for privacy behaviors (Lines \ref{line:url} and \ref{line:recipient}). Finally, \daint's results are sent back (Line \ref{line:result}) to the user as a Push Notification. 

\begin{algorithm}[t!]
\footnotesize
 \caption{\textit{Sink-call recipient analysis and privacy classification of IoT strings at runtime}}
 \label{algo:analytics}
 \begin{algorithmic}[1]

\STATE $inputs \gets$ \daint's API content                         \label{line:api}
\STATE $userinfo \gets input.userinfo$                             \label{line:inistart}  
\STATE $labels \gets input.selectedLabels$
\STATE $content \gets input.content$
\STATE $recipient \gets input.recipients$
\STATE $url \gets input.url$
\STATE $flowResult \gets$ NULL  
\STATE $labelResults \gets$ NULL
\STATE $semanticsLabels \gets$ NULL\\                              \label{line:iniend}

\FOR{$input \in inputs$} 
    \IF{$(content \neq$ NULL)}
        \STATE{$labelResult \gets$ NLP($content$)}                 \label{line:nlp}
    \ENDIF
    \IF{$url$ is insecure}                                        \label{line:url}
        \STATE{$flowResult \gets$ "privacy concern"}
    \ENDIF
    \IF{$recipient \notin user-info$}                                 \label{line:recipient}
        \STATE{$flowResult \gets$ "privacy concern"}
    \ENDIF
\ENDFOR
\STATE sendResults($flowResult$, $labelsList$)       \label{line:result}
\end{algorithmic}
\end{algorithm}

\subsection{More Details on Model Construction}
\label{sec:initialmodel}
Our search for an adequate corpus that characterizes IoT app's data-flows faced particular challenges. First, we could not find any existing IoT corpus. Second, most of the datasets available online contain raw unlabeled data that would require a considerable amount of pre-processing time and resources. Third, the privacy labels considered by \daint could not be inferred from n-gram shingles extracted from a single corpus only. Thus, we combined different knowledge-based datasets to create a larger corpus. We combined the natural language datasets from Google Books N-grams~\cite{google} and Wikidata~\cite{wikipedia}, which contain strings related to geographic (location), economic (devices, user's goods), climate (location), and encyclopedic (general knowledge) datasets. We structured, cleaned, and manually labeled the crawled data. First, we divided the corpora into single shingles (i.e., n-grams of n=1). Then, for cleaning purposes, we filtered out punctuation and stop words. We tested the first NLP model with 61 IoT strings extracted from 45 market IoT apps~\cite{Official, Community}. An average value of 72\% accuracy proved that the first considered model could not accurately represent information extracted from IoT environments. Based on these results, we decided to train the NLP model using specific IoT corpora only.

\begin{table}[!t]
\centering
\setlength{\tabcolsep}{2.5pt}
\resizebox{\columnwidth}{!}{
\scriptsize
\begin{tabular}{ccccccc}
\toprule
\textbf{\begin{tabular}[c]{@{}c@{}}Label \\ Category\end{tabular}} & 
\textbf{\begin{tabular}[c]{@{}c@{}}Messages\\ without \\Leaks or \\Privacy \\Concerns\end{tabular}} & 
\textbf{\begin{tabular}[c]{@{}c@{}}Internet\\ without \\Leaks or \\Privacy \\Concerns\end{tabular}} & 
\textbf{\begin{tabular}[c]{@{}c@{}}Messages\\ with \\Leaks or \\Privacy \\Concerns\end{tabular}} & 
\textbf{\begin{tabular}[c]{@{}c@{}}Internet\\ with \\Leaks or \\Privacy \\Concerns\end{tabular}}  & 
\textbf{\begin{tabular}[c]{@{}c@{}}Total\end{tabular}} & 
\textbf{\begin{tabular}[c]{@{}c@{}}\% Total\end{tabular}} \\ \hline
\midrule
\rowcolor{light-gray}
\textbf{Device-info}    & 47 & 12 & 43 & 17 & 119 & 52.3 \\ 
\textbf{Date-time}  & 7 & 1  & 7  & 0 & 15  & 6.6 \\
\rowcolor{light-gray}
\textbf{User-behavior}      & 21 & 2 & 34 & 10 & 67 & 29.3 \\ 
\textbf{Location}  & 8 & 0 & 14 & 5 & 27 & 11.8 \\ \hline
\rowcolor{light-gray}
\textbf{Total}  & 83 & 15 & 98 & 32 & 228 & 100 \\
\bottomrule
\end{tabular}
}
\caption{Distribution of privacy labels used during \daint's evaluation.}
\vspace{-0.2in}
\label{tab:labeling}
\end{table}

\begin{table*}[t!]
\centering
\normalsize
\begin{tabular}{ccccccccccc}
\toprule
\multirow{ 2}{*}{\textbf{\begin{tabular}[c]{@{}c@{}}Evaluation \\ Metric\end{tabular}}} &
\multicolumn{9}{c}{\textbf{Classification Thresholds}}\\ &
\textbf{\begin{tabular}[c]{@{}c@{}}0.1\end{tabular}} & \textbf{\begin{tabular}[c]{@{}c@{}}0.2\end{tabular}} & \textbf{\begin{tabular}[c]{@{}c@{}}0.3\end{tabular}} & 
\textbf{\begin{tabular}[c]{@{}c@{}}0.4\end{tabular}} & 
\textbf{\begin{tabular}[c]{@{}c@{}}0.5\end{tabular}} & 
\textbf{\begin{tabular}[c]{@{}c@{}}0.6\end{tabular}} & 
\textbf{\begin{tabular}[c]{@{}c@{}}0.7\end{tabular}} & 
\textbf{\begin{tabular}[c]{@{}c@{}}0.8\end{tabular}} & 
\textbf{\begin{tabular}[c]{@{}c@{}}0.9\end{tabular}} &  
\textbf{\begin{tabular}[c]{@{}c@{}}Average\end{tabular}} \\\hline
\midrule
\rowcolor{light-gray}
\textbf{Accuracy}      & 0.9537  & 0.9623  & 0.9692  & 0.9743  & 0.9623  & 0.9469  & 0.9296  &   0.9092    & 0.8750 &  0.9425  \\ 
\textbf{Recall}        & 0.9754  & 0.9400  & 0.9379  & 09341  & 0.8821  & 0.8419  &   0.7919  &   0.7359    & 0.6263 &  0.8517  \\ 
\rowcolor{light-gray}
\textbf{Precision}     & 0.8470  & 0.8776  & 0.9163  & 0.9587  & 0.9803  & 0.9833  &  0.9929 & 0.9975    & 0.9835    & 0.9501  \\  
\textbf{Specificity}   & 0.9283  & 0.9498  & 0.9682  & 0.9772  & 0.9823  & 0.9855  &  0.9876  & 0.9907      & 0.9907  &  0.9734\\ 
\bottomrule
\end{tabular}
\caption{Evaluation metric results in classifying IoT strings for all classification thresholds. The rightmost column presents the average metrics.}
\vspace{-0.1in}
\label{tab:average}
\end{table*}

\begin{table*}[!t]
\centering
\normalsize
\begin{tabular}{cccccccccc}
\toprule
\textbf{\begin{tabular}[c]{@{}c@{}}Privacy\\Label\end{tabular}} & 
\textbf{\begin{tabular}[c]{@{}c@{}}No. of Times\\ Evaluated\end{tabular}} & 
\textbf{\begin{tabular}[c]{@{}c@{}}TP\end{tabular}} & 
\textbf{\begin{tabular}[c]{@{}c@{}}TN\end{tabular}}  & 
\textbf{\begin{tabular}[c]{@{}c@{}}FP\end{tabular}} & 
\textbf{\begin{tabular}[c]{@{}c@{}}FN\end{tabular}} & 
\textbf{\begin{tabular}[c]{@{}c@{}}Accuracy\end{tabular}} & 
\textbf{\begin{tabular}[c]{@{}c@{}}Recall\end{tabular}} & 
\textbf{\begin{tabular}[c]{@{}c@{}}Precision\end{tabular}} & 
\textbf{\begin{tabular}[c]{@{}c@{}}Specificity\end{tabular}}\\ \hline
\midrule
\rowcolor{light-gray}
\textbf{Device-info}    & 1285 & 958 & 232   & 11 & 84  & 0.9260 & 0.9214 & 0.9890 & 0.9547\\ 
\textbf{Date-time}  & 1314 & 99  & 1161  & 18 & 36  & 0.9589 & 0.7333 & 0.8918 & 0.9847\\
\rowcolor{light-gray}
\textbf{User-behavior}      & 1313 & 508 & 684   & 26 & 95  & 0.9079 & 0.8425 & 0.9585 & 0.9633\\ 
\textbf{Location}  & 1314 & 221 & 1061  & 10 & 22  & 0.9756 & 0.9095 & 0.9610 & 0.9907\\
\bottomrule
\end{tabular}
\caption{Evaluation results of \daintf in classifying IoT strings to all the different privacy labels.}
\vspace{-0.2in}
\label{tab:exfiltrations}
\end{table*}

\subsection{Assigning Privacy Labels}
\label{sec:labels}
Table \ref{tab:labeling} summarizes the distribution of the privacy labels during evaluation. One can notice that, out of 228 different labels assigned, 52.3\% corresponds to device-info, followed by a 29.3\% of user-behavior information. We also show in Table \ref{tab:labeling} the number of labels assigned to the different call types (i.e., messaging and Internet). For instance, 83 labels were assigned to strings extracted from messaging that do not leak information, while 98 labels were assigned to strings extracted from messaging that leaked information. On the other hand, Internet communications without privacy concerns received 15 privacy labels while 32 were assigned to Internet communications that leak data. With this distribution, one can notice that the majority of labels were assigned to leaked data. These results reflect on the fact that privacy leakage constitute a serious problem in IoT. %Indeed, the more privacy-sensitive labels assigned to potential leaks, the higher the risk of users' and system's sensitive information being leaked by IoT apps. 

\section{Additional Details on \daint API} \label{sec:apiappendix}
Listing \ref{json} illustrates an example of a JSON object used to send data from an IoT app to \daint analyzer. Once the app data is received, the analyzer extracts the data required to enable recipient analysis and NLP-based classification of the IoT strings. The API also handles \daint's privacy notifications to the user. Once the privacy analysis is completed, \daint sends back to the user another JSON object containing its findings (Listing \ref{jsonback}). We implemented the API using the \textit{asynchttpv1} class of Samsung SmartThings~\cite{smartThings-documentation} which allows asynchronous and encrypted data exchange between the instrumentor and the analyzer.

\begin{lstlisting}[float=t!, caption= An example of a JSON object sent from an IoT app to \daint for further analysis., label=json, escapechar=!, language=html]
/* An example of a JSON object sent to Daint analytics tool */ 

 data "{
  'exfiltration':{
    'texttype':'PLAIN_TEXT',
    'calltype':'Messaging',
    'phone':'111-111-1111',
    'content':'The door was opened for 10 min'
    'userrecipients':'123-456-7890',
  }
}" "https://iotwatchanalyticstool.com/classifytext/"
\end{lstlisting}

\begin{lstlisting}[float=t!, caption= An example \daint response as a JSON object received by an IoT app. ,label=jsonback, escapechar=!, language=html]
/* An example of a JSON object as response from our analytics tool */ 

 data "{
  'exfiltration':{
    'texttype':'PLAIN_TEXT',
    'classification':['device-info', 'date-time']',
    'risklevel': 'privacy concern'
  }
}"
\end{lstlisting}
\vspace{-0.3cm}
%\begin{comment}
\section{Evaluation Metrics}
\label{sec:metrics}
The performance metrics used during \daint's evaluation:

\begin{enumerate}[noitemsep,topsep=0pt]
    \item \textit{True Positive (TP)} represents the number of times a privacy label is correctly applied to an IoT string for certain threshold $th$.
    \item \textit{True negative (TN)} represents the number of times a privacy label is correctly discriminated for certain threshold $th$.
    \item \textit{False Positive (FP)} is the number of times a privacy label is incorrectly assigned to certain IoT string for certain threshold $th$.
    \item \textit{False Negative (FN)} is the number of times a privacy label is incorrectly discriminated for certain threshold $th$.
    \item \textit{Accuracy} is the overall ability of \daint to correctly apply privacy labels to the IoT string for every different $th$.
    \item \textit{Recall} is the ability of the classifier to correctly assign the privacy labels to a specific IoT string after considering both the correctly classified and the incorrectly ignored privacy labels for every value of $th$.
    \item \textit{Precision} is the ability of our classifier to correctly apply the privacy labels to a specific IoT string after considering both the correct and the incorrectly applied privacy labels for every value of $th$.
    \item \textit{Specificity} is the ability of our tool to discriminate the privacy labels for every different $th$.
\end{enumerate}

The performance metrics are defined by the following equations:
\begin{equation}\label{eq:accuracy}
Accuracy = \frac{(T_P+T_N)}{(T_P+T_N+F_P+F_N)},
\end{equation}
\begin{equation}\label{eq:recall}
Recall = \frac{T_P}{(T_P+F_N)},
\end{equation}
\begin{equation}\label{eq:precision}
Precision = \frac{T_P}{(T_P+F_P)},
\end{equation}
\begin{equation}\label{eq:specificity}
Specificity = \frac{T_N}{(T_N+F_P)}.
\end{equation}
%\end{comment}
\vspace{-0.2in}
\section{Additional Evaluation Details}
\label{sec:threshold}

Table \ref{tab:average} details average metric values in classifying IoT strings to privacy labels for every different $th$. Also, we present the average metric values after combining results from all considered privacy labels and thresholds. One can verify that \daint obtains the best performance for threshold values of 0.4, which supports the results showed in Section \ref{sec:evaluation}, Figure \ref{fig:ave_metrics}. For $th$ values higher than 0.5, the accuracy decreases for the privacy labels of device-info and user-behavior. This is mainly because the evaluation of semantically-limited strings related to these two privacy labels requires more sophisticated analysis. For instance, user-behavior achieved incorrect or lower classification scores as the string ``IoT switched to sleep mode'' cannot be easily related to user activities. We obtained similar results for recall metrics; however, recall values of date-time are lower compared to other labels. We found that date-time information could be easily missed from short strings, which makes the label specially vulnerable to false negative events. The precision and specificity improves with $th$ for all the labels, which denotes a remarkable confidence of the model for those results with the highest classification scores. Finally, Table~\ref{tab:exfiltrations} illustrates the performance of \daint in classifying IoT strings for every different privacy label. 

\end{document}